%% file: ms.tex
\documentclass[iop,apj]{emulateapj}
\usepackage{graphics,graphicx,subfigure,apjfonts}

\def\cxo{{\em Chandra}}
\def\xmm{{\em XMM}}
\def\asca{{\em ASCA}}
\def\chic{ChIcAGO}

\def\erg{erg~cm$^{-2}$~s$^{-1}$}

\def\etal{et al.}

\begin{document}

\title{Multi-wavelength Observations of the Radio Magnetar PSR~J1622--4950 and Discovery of its Possibly Associated Supernova Remnant}
\shorttitle{Transient Magnetar PSR~J1622--4950 and its SNR}
\shortauthors{G. E. Anderson et al.}
\submitted{Accepted to {\em The Astrophysical Journal} 2012 March 9}
\author{Gemma E. Anderson\altaffilmark{1}, 
B. M. Gaensler\altaffilmark{1}, 
Patrick O. Slane\altaffilmark{2}, 
Nanda Rea\altaffilmark{3}, 
David L. Kaplan\altaffilmark{4}, 
Bettina Posselt\altaffilmark{5}, 
Lina Levin\altaffilmark{6,7}, 
Simon Johnston\altaffilmark{7}, 
Stephen S. Murray\altaffilmark{8},
Crystal L. Brogan\altaffilmark{9},
Matthew Bailes\altaffilmark{6},
Samuel Bates\altaffilmark{10,11},
Robert A. Benjamin\altaffilmark{12}, 
N.D. Ramesh Bhat\altaffilmark{6},
Marta Burgay\altaffilmark{13},
Sarah Burke-Spolaor\altaffilmark{7,14},
Deepto Chakrabarty\altaffilmark{15},
Nichi D'Amico\altaffilmark{13}, 
Jeremy J. Drake\altaffilmark{2}, 
Paolo Esposito\altaffilmark{13},
Jonathan E. Grindlay\altaffilmark{2}, 
Jaesub Hong\altaffilmark{2}, 
G. L. Israel\altaffilmark{16},
Michael J. Keith\altaffilmark{7},
Michael Kramer\altaffilmark{10,17},
T. Joseph W. Lazio\altaffilmark{14}, 
Julia C. Lee\altaffilmark{2}, 
Jon C. Mauerhan\altaffilmark{18},
Sabrina Milia\altaffilmark{13,19},
Andrea Possenti\altaffilmark{13},
Ben Stappers\altaffilmark{10},
Danny T. H. Steeghs\altaffilmark{20}
}

\altaffiltext{1}{Sydney Institute for Astronomy, School of Physics A29, The University of Sydney, NSW 2006, Australia: g.anderson@physics.usyd.edu.au}
\altaffiltext{2}{Harvard-Smithsonian Center for Astrophysics, Cambridge, MA 02138, USA}
\altaffiltext{3}{Institut de Ciencies de l'Espai (CSIC-IEEC), Campus UAB, Facultat de Ciencies, Torre C5-parell, 2a planta, 08193, Bellaterra (Barcelona), Spain}
\altaffiltext{4}{Department of Physics, University of Wisconsin, Milwaukee, WI 53201, USA}
\altaffiltext{5}{Department of Astronomy and Astrophysics, Pennsylvania State University, PA 16802, USA}
\altaffiltext{6}{Centre for Astrophysics and Supercomputing, Swinburne University of Technology, VIC 3122, Australia}
\altaffiltext{7}{Australia Telescope National Facility, CSIRO Astronomy and Space Science, P.O. Box 76, Epping, NSW 1710, Australia}
\altaffiltext{8}{Department of Physics and Astronomy, John Hopkins University, Baltimore, MD 21218, USA}
\altaffiltext{9}{National Radio Astronomy Observatory, Charlottesville, VA 22903, USA}
\altaffiltext{10}{Jodrell Bank Centre for Astrophysics, School of Physics and Astronomy, University of Manchester, Manchester M13 9PL, UK}
\altaffiltext{11}{Physics Department, West Virginia University, Morgantown WV, 2650}
\altaffiltext{12}{Department of Physics, University of Wisconsin, Whitewater, WI 53190, USA}
\altaffiltext{13}{INAF/Osservatorio Astronomico di Cagliari, 09012 Capoterra, Italy}
\altaffiltext{14}{Jet Propulsion Laboratory, California Institute of Technology, Pasadena, CA 91109, USA}
\altaffiltext{15}{MIT Kavli Institute for Astrophysics and Space Research and Department of Physics, Massachusetts Institute of Technology, Cambridge, MA 02139, USA}
\altaffiltext{16}{INAF/Osservatorio Astronomico di Roma, Monteporsio Catone, Italy}
\altaffiltext{17}{Max Planck Institut f\"{u}r Radioastronomie, 53121 Bonn, Germany}
\altaffiltext{18}{Spitzer Science Center, California Institute of Technology, Pasadena, CA 91125, USA}
\altaffiltext{19}{Dipartimento di Fisica, Universita' degli Studi di Cagliari, Cittadella Universitaria, 09042 Monserrato (CA), Italy}
\altaffiltext{20}{Department of Physics, University of Warwick, Coventry CV4 7AL, UK}

\begin{abstract}

We present multi-wavelength observations of the radio magnetar PSR~J1622--4950 and its environment. Observations of PSR J1622--4950 with \cxo\ (in 2007 and 2009) and \xmm\ (in 2011) show that the X-ray flux of PSR J1622--4950 has decreased by a factor of $\sim50$ over 3.7 years, decaying exponentially with a characteristic time of $\tau = 360 \pm 11$ days. This behavior identifies PSR J1622--4950 as a possible addition to the small class of transient magnetars. The X-ray decay likely indicates that PSR~J1622--4950 is recovering from an X-ray outburst that occurred earlier in 2007, before the 2007 \cxo\ observations. Observations with the Australia Telescope Compact Array show strong radio variability, including a possible radio flaring event at least one and a half years after the 2007 X-ray outburst that may be a direct result of this X-ray event. Radio observations with the Molonglo Observatory Synthesis Telescope reveal that PSR J1622--4950 is $8'$ southeast of a diffuse radio arc, G333.9+0.0, which appears non-thermal in nature and which could possibly be a previously undiscovered supernova remnant. If G333.9+0.0 is a supernova remnant then the estimates of its size and age, combined with the close proximity and reasonable implied velocity of PSR J1622--4950, suggests that these two objects could be physically associated.

\end{abstract}

\keywords{ISM: individual(G333.9+0.0) -- pulsars: individual (PSR~J1622--4950) -- radio continuum: stars -- stars: neutron -- supernova remnants -- X-rays: stars}

\section{Introduction}

``Magnetar'' has become the commonly used term to describe the emerging class of rare, young and highly magnetized neutron stars (B $\gtrsim 10^{14}$ G) referred to as Anomalous X-ray Pulsars (AXPs) and Soft Gamma Repeaters (SGRs), with perhaps as many as 23 now detected.\footnote{http://www.physics.mcgill.ca/$\sim$pulsar/magnetar/main.html} Magnetars are primarily bright X-ray emitters, for which most of this high energy radiation is thought to be generated through the decay of their strong magnetic fields \citep[for a review on magnetars see][]{mereghetti08}. While magnetars may be X-ray luminous, only three such sources have been detected at radio wavelengths: \object[XTE J1810-197]{XTE J1810--197} \citep{halpern05,camilo06Nat}, \object[1E 1547.0-5408]{1E 1547.0--5408} \citep{camilo07v666} and most recently \object[PSR J1622-4950]{PSR J1622--4950} \citep{levin10}.

PSR J1622--4950 is unusual as it is the first magnetar to have been discovered by its pulsed radio emission \citep{levin10}. PSR J1622--4950 was detected as a 4.3s period radio pulsar in the High Time Resolution Universe survey performed with the Parkes 64m telescope \citep{keith10}, and was then recovered in other archival radio data-sets \citep{levin10}. \citet{levin10} showed this pulsar to be very different from ordinary radio pulsars in that it displays large variations in flux over time, with time spans of inactivity during which it is undetected. Its variable pulse profile and high inferred magnetic field strength show that it has similar properties to the only other two known radio magnetars, XTE J1810--197 and 1E 1547.0--5408. Both of these radio magnetars also fit into an emerging subgroup known as transient magnetars, which randomly undergo bright X-ray bursts with increases in X-ray flux up to a factor of several 100 \citep[for example see][]{tam06,bernardini09,bernardini11,scholz11}. It appears that for both XTE J1810--197 and 1E 1547.0--5408, the pulsed radio emission turned on as a result of an X-ray outburst.

In this paper we present multi-wavelength data that further confirm the magnetar identification of PSR J1622--4950 and that support its transient nature. New and archival radio, infrared and X-ray observations are presented in Sections~2 and the corresponding results are presented in Section~3. In Section~4 we use these data to explore PSR J1622--4950's X-ray and radio variability over the last ten years, and to study the variable nature of its polarized radio emission. We also discuss a possible $\gamma$-ray counterpart to PSR J1622--4950, and identify a possibly associated young supernova remnant (SNR).

\section{Observations}

\subsection{X-ray Observations}

\subsubsection{\cxo\ Observations}

The position of PSR J1622--4950 was observed three times with the \textit{Chandra X-ray Observatory}, using the Advanced CCD Imaging Spectrometer \citep[ACIS;][]{garmire03}. The first observation took place on 2007 June 13 as part of the ``ChIcAGO'' (Chasing the Identification of \asca\ Galactic Objects) project. \chic\ is a survey designed to localize and classify the unidentified X-ray sources discovered during the \asca\ Galactic Plane Survey (AGPS) \citep[see][for some of the first ChIcAGO results; further details will be published by G. E. Anderson et al., in preparation.]{anderson11} PSR J1622--4950 was observed again on 2009 June 14 and on 2009 July 10. These data were reduced using the Chandra Interactive Analysis of Observations (CIAO) software, version 4.3, following the online CIAO 4.3 Science Threads\footnote{http://cxc.harvard.edu/ciao/threads/}. For details on these \cxo\ observations see Table~\ref{tab:1}.

\subsubsection{\textit{XMM-Newton} Observations}

We observed PSR J1622--4950 with {\em XMM--Newton} starting on 2011 Feb 22. The PN \citep{struder01} and MOS \citep{turner01} cameras, which comprise the European Photon Imaging Camera (EPIC), were operated in Full Frame mode. Data were processed using {\tt SAS} version 10.0.0,\footnote{http://xmm.esac.esa.int/sas/} and we employed the latest calibration files available at the time the reduction was performed (2011 March). Standard data-screening criteria were applied in the extraction of scientific products. The observation was highly affected by proton flares, which we had to cut from our data before proceeding with the scientific analysis, resulting in a net livetime exposure of 46.4 ks. Since both MOS and PN give consistent results, in the following we report only on the high-time-resolution PN data. Further details on the \xmm\ EPIC-PN observation can be found in Table~\ref{tab:1}.

\subsection{Infrared and Optical Observations}

We compared the \cxo\ position of PSR J1622--4950 (see Section 3.1.1. below) to optical and infrared point source catalogs and survey images including the US Naval Observatory (USNO) B catalog, version 1.0 \citep{monet03}, the Two Micron All Sky Survey \citep[2MASS;][]{skrutskie06}, the Galactic Legacy Infrared Mid-Plane Survey Extraordinaire \citep[GLIMPSE;][]{benjamin03}, and the 24 {\micron} images from the MIPSGAL Survey \citep{carey09} but no counterpart to PSR J1622--4950 was identified in any of these data sets. On 2007 June 24 we observed the position of PSR J1622--4950 in the $K_{s}$ band for 13.5 minutes using the Persson's Auxiliary Nasmyth Infrared Camera \citep[PANIC:][]{martini04,osip08} on the 6.5m Magellan I (Baade) telescope, located at Las Campanas Observatory. The final image was a combination of 30s exposures carried out three times at each position of a nine-point dither pattern to account for the high background. Standard reduction procedures were applied using the IRAF software package \citep{tody86,tody93}. We calibrated the field's photometry and astrometry using the source extraction software SExtractor \citep{bertain96} with comparisons to the Two Micron All Sky Survey point source catalog (2MASS PSC), which has a position uncertainty of $0.''1$ \citep{skrutskie06}. No infrared counterpart was detected within $0.''8$ of the X-ray position of PSR J1622--4950 to a lower limit $K_{s} \geq 20.7$. 

\subsection{Radio Observations}

\subsubsection{Archival}

PSR J1622--4950 is coincident with a bright knot of emission that is part of a larger diffuse radio source seen in the first and second epoch Molonglo Galactic Plane Surveys \citep[MGPS1 and MGPS2 respectively;][]{green99,murphy07} and also seen, albeit less clearly, in the continuum maps from the Southern Galactic Plane Survey \citep[SGPS:][]{haverkorn06}. This feature will be discussed in Section 3.2.1. The MGPS surveys were performed at 843 MHz with the Molonglo Observatory Synthesis Telescope (MOST) at a resolution of $43''$ while the continuum SGPS images were created from Australia Telescope Compact Array (ATCA) 1.4 GHz observations at a resolution of $100''$. 

\subsubsection{ATCA Observations}

We observed PSR J1622--4950 with the ATCA simultaneously at both 5 and 9 GHz, with a 128 MHz bandwidth in each band, on 2008 Nov 22 and 2008 Dec 5 as summarized in Table~\ref{tab:2}. A mosaic pattern was used in an attempt to both detect the magnetar and study the morphology of the surrounding diffuse emission seen in the MGPS1, MGPS2 and SGPS. Both frequencies were observed for 6.79 and 8.01 hours in the EW367 and 750B configurations, respectively, using \object[PKS B1934-638]{PKS B1934--638} for flux calibration and \object[PMN J1603-4904]{PMN J1603--4904} for phase and polarization calibration. Data reduction and analysis were performed using the MIRIAD\footnote{http://www.atnf.csiro.au/computing/software/miriad/} software package using standard techniques. 

\section{Results}

\subsection{X-ray Results}

\subsubsection{X-ray Counterpart to PSR J1622--4950}

The 2007 June \cxo\ observation was an attempt to localize the unidentified AGPS source \object[AX J1622.7-4946]{AX J162246--4946} as part of the \chic\ survey. Six point sources were detected within $4'$ of the position of AX J162246--4946, all of which may have contributed to the X-ray emission originally detected with \asca\ on 1999 April 22 as part of the AGPS. The brightest source, \object[CXOU J162244.8-495054]{CXOU J162244.8--495054} \citep{evans10}, which \citet{levin10} identified as the X-ray counterpart to PSR J1622--4950, was at least 20 times brighter than the other five sources. As CXOU J162244.8--495054 fell in the gap between CCDs for part of the 2007 June observation we only included those time intervals where the source was detected, resulting in an effective exposure time of $\sim2.02$ ks and a count-rate of $96.8 \pm 6.9$ counts ks$^{-1}$ in the $0.3-8.0$ keV energy range. 

The 2009 July \cxo\ observation, two years later, also detected the same six sources in the $4'$ region surrounding the position of AX J162246--4946. While the 2009 July observation was $\sim 10$ times longer than the 2007 June observation, CXOU J162244.8--495054 had faded significantly with a much lower count-rate of $7.1 \pm 0.6$ counts ks$^{-1}$ (equivalent to $\sim9.3$ counts ks$^{-1}$ when compared to the 2007 June observation in the ACIS-S configuration\footnote{All count-rate conversions are based on X-ray spectral fit of the 2007 June \cxo\ observation, derived in Section 3.1.2., in order to account for the differences in the off-axis position and between the X-ray telescope instruments.}) in the $0.3-8.0$ keV energy range. The X-ray flux of CXOU J162244.8--495054 had therefore reduced by a factor of $\sim10$ in the two years between the 2007 June and 2009 July \cxo\ observations. The best X-ray position for CXOU J162244.8--495054, to within a $0''.8$ radius circle for a 95\% confidence,\footnote{This error takes into account the \cxo\ absolute positional accuracy (see http://cxc.harvard.edu/cal/ASPECT/celmon/) and the position uncertainties associated with the CIAO source detection algorithm \texttt{wavdetect} \citep{hong05}.} is (J2000) RA = 16:22:44.89 and Dec = -49:50:52.7 taken from the 2009 July \cxo\ observation. This agrees within 95\% confidence of the X-ray position given by \citet{levin10}. From now on we will refer to both the X-ray counterpart, CXOU J162244.8--495054, and radio counterpart of this magnetar as PSR J1622--4950.  

 The \cxo\ observation from approximately one month earlier (2009 June) also detected PSR J1622--4950 very far off-axis, showing a count-rate slightly higher than that of the 2009 July observation. This is equivalent to $\sim10.8$ counts ks$^{-1}$ in the $0.3-8.0$ keV energy range when compared to the 2007 June \cxo\ observation. In 2011 Feb, our \xmm\ observation showed that PSR J1622--4950 had faded even more, resulting in an EPIC-PN count-rate of $5.8\pm0.6$ count ks$^{-1}$ in the $0.4-10.0$ keV energy range (equivalent to $\sim1.8$ counts ks$^{-1}$ in the ACIS-S configuration of 2007 June in the $0.3-8.0$ keV energy range). A summary of each X-ray observation can be found in Table~\ref{tab:1}.

It is also worth exploring to what extent the X-ray emission from PSR J1622--4950 may have contributed to the flux from AX J162246--4946 in the 1999 \asca\ observation. PSR J1622--4950 lies $4'$ away from the position of AX J162246--4946 \citep{sugizaki01}. The other 5 point sources that were detected in both the 2007 June and 2009 July \cxo\ observations all fall within $3'$ of the position of AX J162246--4946. By fitting an absorbed power law to the spectrum of each of these point sources, assuming a power law spectral index of 2 and $N_{H} = 2 \times 10^{22}$ cm$^{-2}$, we estimate they contributed a total absorbed flux of $\sim1 \times 10^{-13}$ \erg in the $0.3-10.0$ keV energy range. (The fluxes of 3 out of the 5 point source were unchanged within a factor of 2 between the 2007 June and 2009 July \cxo\ observations. The flux of the fourth point sources may have decreased by a factor of $\sim4$ between 2007 and 2009 and the flux of the fifth point source may have increased by a factor of $\sim6$.) Using the $2.0-10.0$ keV \asca\ count-rates \citep{sugizaki01} and the absorbed power law spectral fit described above, we estimate an absorbed X-ray flux of $\sim4 \times 10^{-13}$ \erg\ from AX J162246--4946. Subtracting the contribution from the 5 nearby point sources described above, we estimate that the unabsorbed X-ray flux from PSR J1622--4950 contributed, at most, $\sim3 \times 10^{-13}$ \erg\ (corresponding to 75\%) to the observed X-ray emission seen from AX J162246-4946 in the AGPS.

\subsubsection{X-ray Spectrum and Variability}

In all three \cxo\ observations we extracted the X-ray spectrum of PSR J1622--4950 using the CIAO tool \texttt{specextract}. As the source fell close to the edge of the CCD in the 2007 June observation and was very far off-axis in the 2009 June observation, \texttt{specextract} was run in extended source mode to better handle the creation of response files. The spectra from the 2007 June and 2009 July \cxo\ observations were binned before fitting to include at least 10 counts per bin, as this is the minimum number of counts required to result in a statistically significant fit. The small number of counts detected, combined with the large absorption in the Galactic plane, resulted in very few photons being detected below 2 keV in these two \cxo\ observations. A larger number of counts were detected during the 2009 June \cxo\ observation (the source had faded but the observation was long with the source off-axis) so we chose to bin its spectrum to include at least 40 counts per bin. The \xmm\ EPIC-PN source events and spectrum were extracted within a circular region of 20$^{\prime\prime}$ centered on the peak of the point spread function of the source, while the background was obtained from a source-free region of similar size. When generating the EPIC-PN spectrum we have included events with PATTERN $\leq4$ (i.e. single and double events). This \xmm\ spectrum was binned before fitting, using at least 40 counts per bin to compensate for the high background and not oversampling the intrinsic spectral resolution by more than a factor of 3.\footnote{http://xmm.esa.int/sas/10.0.0/doc/specgroup/node14.html}

The X-ray spectra of magnetars are commonly fit using a blackbody plus power-law model or a multiple blackbody model. We chose to simultaneously fit the four X-ray observations of PSR J1622--4950 with both a single absorbed blackbody model and a single absorbed power-law model as the low number of counts preclude the identification of multiple components. The spectra were fit using \texttt{XSPEC} \citep{dorman01} with the absorption, $N_{H}$, set as a free parameter but constrained to have the same value at each epoch, while the other parameters were set to vary individually. It should be noted that the resulting spectrum and fit to the June 2007 \cxo\ observation of PSR J1622--4950 may not be entirely accurate given its location close to the edge of the CCD, which resulted in the source dithering off the chip at regular intervals. By removing those time intervals where the source was not on the chip we have reduced this error but there can still be problems associated with the response files created during the extraction process.\footnote{For caveats associated with the analysis of sources near the edge of a CCD see http://cxc.harvard.edu/ciao/threads/specextract/index.html\#caveat.acisedge and http://cxc.harvard.edu/cgi-gen/wonderdesk/wonderdesk.cgi?do=faq\_view \_record\&faq=1\&view\_detail=1\&faq\_id=66}

A simultaneous spectral fit with either an absorbed blackbody or an absorbed power-law model are equally reasonable for describing the four X-ray spectra of PSR J1622--4950, both resulting in $\chi^{2}_{red} = 0.7$. The blackbody and power-law spectral fits resulted in absorptions of $N_{H} = 5.4^{+1.6}_{-1.4} \times 10^{22}$ cm$^{-2}$ and $N_{H} = 10.5^{+2.5}_{-2.1} \times 10^{22}$ cm$^{-2}$ (errors at 90\% confidence), respectively, using abundances from \citet{lodders03} and the photoelectric scattering section from \citet{balucinska92} and \citet{yan98}. These absorptions exceed the Galactic $N_{H}$ at the position of PSR J1622–-4950 predicted from H{\sc i} surveys \citep{dickey90,kalberla05} by a factor of 2-4, as expected at high column densities and low Galactic latitudes \citep[e.g.][]{arabadjis99}. The absorbed blackbody and absorbed power-law fit parameters for each spectrum are listed in Table~\ref{tab:3}. 

In order to investigate the long-term X-ray variability of PSR J1622--4950 we computed the absorbed and unabsorbed fluxes and their uncertainties for the four X-ray spectra, using the \texttt{XSPEC} model \texttt{cflux}, in the $0.3-10.0$ keV energy range. These fluxes are listed in Table~\ref{tab:3}. (As the absorbed X-ray fluxes calculated from the blackbody and power-law spectral fits are the same at each epoch within the 90\% confidence we will use the fluxes obtained from the blackbody fit in the rest of our analysis unless otherwise stated. This investigation was also limited to the absorbed fluxes as the errors on the unabsorbed power law fluxes are very large and unconstraining.) The 2007 June \cxo\ observation is our brightest X-ray detection of PSR J1622--4950, making it 9 times brighter than the two mid 2009 \cxo\ observations and 47 times brighter than in the 2011 Feb \xmm\ observation. As described in Section 3.1.1., PSR J1622--4950 has an X-ray flux upper limit of $F_{x} \leq 3 \times 10^{-13}$ \erg\ ($0.3-10.0$ keV) in the 1999 April \asca\ observation. This flux upper limit is $\gtrsim6$ times fainter than our X-ray flux measurement from the 2007 June \cxo\ observation but is consistent with the fluxes measured at the other three X-ray epochs. 

The best fit blackbody temperatures ($kT$) and power law spectral indices ($\Gamma$) of the simultaneous spectral fits are also listed in Table~\ref{tab:3}. While both of these parameters at each epoch are on the high end when compared to those seen from other known magnetars\footnote{See http://www.physics.mcgill.ca/$\sim$pulsar/magnetar/main.html} they are not unreasonable \citep[for example see][]{gelfand07,enoto09}. The parameters $kT$ and $\Gamma$ also indicate there was no significant spectral variability as these values remain the same between epochs within the 90\% confidence. The spectra with the absorbed blackbodies are plotted in Figure~\ref{fig1}. 

We further investigated evidence of spectral evolution by comparing the hardness ratios resulting from simultaneously fitting all four spectra with an absorbed blackbody model and calculating the flux and 90\% flux error for different energy bands (e.g. 0.3-2 keV, 2-4 keV, etc). The flux hardness ratios for different energy band combinations (soft/hard), propagating through the 90\% errors, shows no evidence of spectral variation between the epochs. While this method is model dependent it also demonstrates there is no strong evolution to PSR J1622--4950's X-ray spectrum.

\citet{levin10} calculated a dispersion measure (DM) distance of $9$~kpc to PSR J1622--4950 using the \citet{cordes02} NE2001 Galactic free electron density model.\footnote{This distance estimate should be treated with caution as proper motion studies have shown that DM-based distance measurements can differ by more than a factor of two to the distance calculated from parallax \citep{chatterjee09,deller09}.} We used this distance to estimate the unabsorbed $0.3-10.0$ keV luminosity at the four epochs from the unabsorbed fluxes calculated from both the blackbody and power-law spectral fits. These values are summarized in Table~\ref{tab:3}. The X-ray luminosities calculated from the blackbody fit show a monotonic reduction over the four epochs and are consistent with the range of luminosities seen from other magnetars.\footnote{See http://www.physics.mcgill.ca/$\sim$pulsar/magnetar/main.html} However, the luminosities calculated from the power-law fit are on the high end of known magnetar luminosities, with the large errors preventing us from observing any obvious evolution between epochs.

\subsubsection{X-ray Timing Analysis}

Only the \xmm\ data were used in our X-ray timing investigation as our \cxo\ ACIS observations do not have sufficient time resolution to detect PSR J1622--4950's pulse period of 4.3s. Using the EPIC-PN observations, which have a time resolution of 73.4 ms, we shifted all photons arrival times to the solar system barycenter, and used the {\texttt XRONOS} software for the timing analysis. We used data in the $0.3-4.0$ keV energy range where the source was relatively bright, then folded the data using the pulsar ephemeris derived from ongoing radio monitoring: P$=4.32645312$ s at Modified Julian Date (MJD) 55586.5 (Levin et al., in preparation). We do not detect a significant signal in the X-ray band with a 3$\sigma$ upper limit on the $0.3-4.0$ keV pulsed fraction of 70\% \citep[derived as by][]{vaughan94,israel96}. The pulsed fraction is defined as $(N_{max}-N_{min})/(N_{max}+N_{min})$, with $N_{max}$ and $N_{min}$ being the maximum and minimum counts of a putative sinusoidal signal at the frequency of PSR~J1622--4950. A similar analysis was also conducted in the $0.3-2.0$ and $2.0-10.0$ keV energy bands as the strength and phase of a magnetar's pulsations can be energy dependent \citep{gotthelf09}. Each band had $<100$ counts with no pulsations detected, resulting in an unconstraining pulse fraction upper limit of $99\%$.

In order to search for short term variability from PSR J1622--4950 in the three \cxo\ observations we corrected the photon arrival times to the solar system barycenter. In the case of the 2007 June observation we also filtered out those times when the source was not on the chip to remove any contribution caused by the source dithering on and off the chip. There was no evidence of short term variability beyond 40\% of the mean count-rate between 250 and 2020s, 5000s and 60870s and 2500s and 20130s in the 2007 June, 2009 June and 2009 July \cxo\ observations, respectively.\footnote{The smaller number in each range is the size of a time bin chosen to include a statistically significant number of counts. The upper number is the exposure time.}

\subsection{Radio Results}

\subsubsection{Archival}

The region surrounding PSR J1622--4950 at radio wavelengths in MGPS1 is shown in Figure~\ref{f2a}. The position of PSR J1622--4950 is denoted by a white ``+'' sign. The diffuse radio emission surrounding PSR J1622--4950, encompassed by the contours seen in Figures~\ref{f2b}~and~\ref{f2c}, is denoted source A. There is a nearby H{\sc ii} region, \object[GAL 333.6-00.2]{G333.6--0.2}, located to the southwest of the pulsar \citep{paladini03}. The other three diffuse radio sources in the immediate vicinity of the pulsar are denoted B, C and D. Figures~\ref{f2b} and \ref{f2c} show the corresponding infrared GLIMPSE 8 {\micron} and MIPSGAL 24 {\micron} views of this area of sky, respectively, overlaid with MGPS 1 as black contours.

The radio source A does not appear to have been catalogued in any surveys of this region. The brightest knot of radio emission in source A is coincident with PSR J1622--4950 and extends out in some directions as far as $4'$. This diffuse radio emission also forms a ring morphology below PSR J1622--4950, which can be seen in Figure~\ref{f2a}. This same morphology was also resolved by \citet{levin10} using the ATCA. Faint diffuse infrared emission is detected all over the immediate field as seen in Figures~\ref{f2b} and \ref{f2c}, but is not concentrated in the vicinity of the bright knot of radio emission in source A. It is therefore more likely that this diffuse infrared emission originates from G333.6--0.2, which is extremely bright in the radio and at 8 and 24 {\micron}. At the Southern edge of source A there is a very bright 24 {\micron} source, \object[IRAS 16190-4946]{IRAS 16190--4946} \citep[ellipse major and minor axis uncertainty of $23''$ and $3''$, respectively;][]{helou88}, indicated by a white $23''$ radius circle in Figure~\ref{f2c}. IRAS 16190--4946 lies $\sim2'$ from PSR J1622--4950 so is unlikely to be associated with source A immediately coincident with the magnetar.

To the northwest of the pulsar there is a radio source in the shape of an arc, labeled B in Figure~\ref{fig2}. Figures~\ref{f2b} and \ref{f2c} show no evidence for diffuse infrared emission matching the structure of source B. If we assume that this arc is part of a circular shell, we measure an equivalent radius of $7' \pm 1'$ with approximate central coordinates of (J2000) RA=16:22:38 and Dec=-49:49:48. The inferred center of the arc is offset $0.8'-2.5'$ from the position of PSR J1622--4950. Another pulsar, \object[PSR J1622-4944]{PSR J1622--4944}, indicated by a black ``+'' sign in Figure~\ref{fig2}, lies $1'$ inside the inner edge of the arc. The distance to PSR J1622--4944 is $\sim7.9$ kpc calculated using the DM published by \citet{manchester01} assuming the \citet{cordes02} model. There is also an infrared dark cloud, SDC G333.900+0.022 \citep{peretto09} coincident with the arc but it is unlikely to be associated given that radio emission is not commonly associated with such clouds and that its morphology does not match that of source B.

A possible counter-arc to source B, labeled source C in Figure~\ref{fig2}, is seen $\sim8'$ to the southeast of PSR J1622--4950 but the lack of symmetry between these arcs renders an association between the two speculative. Radio source D is coincident with the 6.7 GHz maser source G333.93--0.13 \citep{pestalozzi05} and is likely also associated with IRAS 16194--4941. At 24 {\micron} other infrared sources are detected within the brighter radio contours of source D, which are very faint or undetected at 8 {\micron}.

\subsubsection{ATCA Results}

Initial analysis of the 2008 Nov and 2008 Dec ATCA data demonstrated that we did not have enough sensitivity to see the diffuse emission surrounding PSR J1622--4950 as seen in MGPS, SGPS and the \citet{levin10} ATCA observations (which took place on 2009 December 8 and 2010 February 27). While the \citet{levin10} ATCA observations used the same frequencies, array configurations, and integration times as the 2008 ATCA observations presented here, they were using the new ATCA correlator, which has a 2~GHz bandwidth and thus a much higher sensitivity. Regardless of the lack of diffuse emission, our ATCA observations did detect a radio point source at the magnetar's \cxo\ position. We analyzed the 2008 Nov and 2008 Dec observations separately, using just the antenna 6 baselines (i.e., baselines between 5 and 6 km). This significantly reduced sidelobe contamination from G333.6--0.2. 

The resulting time and phase averaged flux densities and polarizations of PSR J1622--4950, measured during the 2008 Nov and 2008 Dec ATCA observations, are listed in Table~\ref{tab:2}. While the 9 GHz flux density remained fairly steady between the two observations, it increased by 22\% at 5 GHz. The ATCA observations of PSR J1622-4950 taken by \citet{levin10} (see Table~\ref{tab:2}) indicate a decrease in radio flux of $\sim68$\% and  $\sim55$\%, at 5 and 9 GHz respectively, approximately one year after our 2008 Dec ATCA observation. 

The flux changes we observed between our 2008 Nov and 2008 Dec ATCA observations demonstrated a steepening in the radio spectral index of PSR J1622--4950 over two weeks, from $\alpha = -0.13 \pm 0.04$ on 2008 Nov 22 to $\alpha = -0.44 \pm 0.04$ on 2008 Dec 5, where $S_{\nu} \propto \nu^{\alpha}$. These values contrast to the positive time-averaged spectral index derived by \citet{levin10}, calculated using a combination of observations with different telescopes in the frequency range $1.4-9$ GHz and over various epochs between 1998 Feb and 2010 Jan. 

The polarization of PSR J1622--4950 also changed significantly between our two ATCA epochs. During the 2008 Nov ATCA observation the linearly polarized fraction was $\sim80$\%, and circular polarization $\lesssim8$\% at both frequencies. Two weeks later the linearly polarized fraction had dropped to $<20\%$  but the source had become $\sim15\%$ circularly polarized. (This circular polarization was negatively handed using the pulsar astronomy sign convention.) It should be noted that these polarization fractions quoted here are lower limits on the true peak polarization of PSR J1622--4950 given that the ATCA observations of the pulse signal are time and phased averaged.

\section{Discussion}

\subsection{Variability}

The spin-down luminosity of PSR J1622--4950 calculated by \citet{levin10} is $\dot{E} = 8.5 \times 10^{33}$ erg s$^{-1}$. Using the unabsorbed blackbody luminosities listed in Table~\ref{tab:3} we find that $L_{x} \sim 3.5 \dot{E}$, $L_{x} \sim 0.6 \dot{E}$, $L_{x} \sim 0.5 \dot{E}$ and $L_{x} \sim 0.1 \dot{E}$ for the 2007 June, 2009 June, 2009 July \cxo\ and 2011 Feb \xmm\ observations, respectively, in the $0.3-10.0$ keV energy range. All these ratios are significantly higher than what we expect from young pulsars but are similar to what we see from other magnetars.\footnote{Typical $\dot{E}$ and $L_{x}$ values for most magnetars are summarized at http://www.physics.mcgill.ca/$\sim$pulsar/magnetar/main.html} One of the defining magnetar characteristics is $L_{x} > \dot{E}$ \citep{mereghetti08} but this relationship can also be the case for cooling neutron stars \citep[for example see][]{vankerkwijk08,halpern11}. However, the X-ray variability and relatively high X-ray luminosity that we observe from PSR J1622--4950 are not consistent with cooling. We therefore confirm the \citet{levin10} identification of PSR J1622--4950 as a magnetar.

The simultaneous blackbody and power-law spectral fits to our four X-ray spectra of PSR J1622--4950 show that over 3.7 years the X-ray flux decreased by a factor of 47. Figure~\ref{fig3} is a light-curve showing the blackbody absorbed X-ray flux values in blue where the \cxo\ observations are denoted by open circles and the \xmm\ observation is a filled circle. Figure~\ref{fig3} clearly illustrates the smooth decay of the X-ray emission from an outburst that may have occurred before or during the 2007 June \cxo\ observation. We therefore suggest that PSR J1622--4950 is a transient magnetar, having demonstrated similar variations in X-ray flux to the two transient magnetars XTE J1810--197 and 1E 1547.0--5408 \citep{gotthelf05,gotthelf07,bernardini09,halpern08}.

The decays of outbursts from transient magnetars have been modeled as exponentials, power laws and multiple power laws \citep[for example see Figure 3 of][]{rea11}. We fitted both an exponential and power-law to the absorbed blackbody X-ray flux values from the \cxo\ and \xmm\ observations. (We determined from these X-ray observations that the cross-calibration between the instruments was no larger than 12\%, which is within the 90\% confidence flux errors.) The decay of the X-ray light-curve is best described by an exponential with a characteristic decay time of $\tau \sim360 \pm 11$ days. This is a similar to the X-ray decay of XTE J1810--197 \citep{bernardini09} but much slower than the decay of the multiple bursts produced by 1E 1547.0--5408 \citep[for example see][]{israel10}.

In Section 3.1.2. we demonstrated that there is no evidence of strong evolution in PSR J1622--4950's X-ray spectrum over the 3.7 years of X-ray observations. A consistent temperature between epochs was also observed from XTE J1810--197, where the blackbody temperature remained unchanged for over three years after the initial outburst \citep{bernardini09}. Further X-ray observations with higher count statistics are required to determine if the X-ray spectrum of PSR J1622--4950 is evolving with time.

If the fading X-ray emission that we have observed from PSR J1622--4950 is thermal, then it is likely emitted by a hot spot on the surface of the magnetar that remains constant in temperature but decreases in size over time. This behavior is potentially explained by the magnetar coronal model \citep{beloborodov07}. Specifically \citet{gotthelf05} suggest that the shrinking of a hot spot after a burst could be due to the decay of currents or rearrangement of the magnetic field lines altering the heat being channeled to the surface. The changes in X-ray flux on the order of weeks to years may be the magnetar's crust plastically responding to the unwinding of these fields, which in turn deposits energy into the magnetosphere resulting in transient behavior \citep[][and references therein]{muno07}.

The radio spectral index of PSR J1622--4950 (listed in Table~\ref{tab:2}) appears to be variable and far flatter than the steep, $\alpha \approx -1.6$, stable average spectral index we expect from ordinary young radio pulsars \citep{lorimer95}. It is, however, very similar in behavior to the other radio magnetars, XTE J1810--197 and 1E 1547.0--5408 \citep{camilo07v669,lazaridis08,camilo08}. The time-variable radio spectral indices of magnetars are not well understood but \citet{thompson08a,thompson08b} suggests that this phenomenon could be the result of current-driven instabilities in the closed magnetosphere of the magnetar. \citet{serylak09} (and references therein) also speculate that the observed flat radio spectra of magnetars could be the result of the open magnetic field lines having a high plasma density.

As mentioned in Section 2.2. our PANIC observations obtained a counterpart lower limit of $K_{s} \geq 20.7$ for PSR J1622--4950. Currently only seven magnetars have been detected in the $K_{s}$-band \citep{rea11}. By correcting for the difference in extinction and relative distances, a $K_{s}$-band counterpart similar to that seen from the majority of these infrared magnetars, at the position of PSR J1622--4950, would be extremely faint, having a magnitude $>24.7$.\footnote{For details on the $K$-band observations and distance estimates of each magnetar see \citet{rea11}, http://www.physics.mcgill.ca/$\sim$pulsar/magnetar/main.html and references therein.} While the variable infrared behavior is not consistent between magnetars the ratio between their X-ray and infrared flux appears to remain fairly consistent where $F_{\nu}\nu \simeq 1\times10^{4}F_{x}$ \citep{durant05}, predicting a $K$-band magnitude of 24.5 for PSR J1622--4950 based on the 2007 June \cxo\ observation. A much deeper infrared observation is therefore required to detect the counterpart to PSR J1622--4950 during a future period of X-ray bursting activity.

\subsection{Polarization}

In Section 3.2.2. we showed that the linearly and circularly polarized fractions of the radio emission from PSR J1622--4950 both changed significantly between 2008 Nov and 2008 Dec. In contrast, phase resolved radio observations of the two radio magnetars XTE J1810--197 and 1E 1547.0--5408 show that their linear polarization is consistently high and that neither their linear or circular polarization exhibit dramatic changes in intensity over time \citep{kramer07,camilo07v659,camilo08}.

The significant linear and circular polarization variability we observed from PSR J1622--4950 can be explained by changes in the geometry of the magnetosphere of the magnetar causing the overall pulse profile to vary \citep{camilo07v663}. The 2008 Nov and 2008 Dec ATCA observations were phase- and time-averaged, amplifying the observed polarization variability over the intrinsic behavior. Results from \citet{levin10} show that, unlike most normal radio pulsars, the radio pulse profile of PSR J1622--4950 is variable between consecutive epochs, similar to the radio pulse profiles of XTE J1810--197 and 1E 1547.0--5408 \citep{camilo07v663,kramer07,camilo08}. In Figure~1 of \citet{levin10}, we see that in three consecutive epochs, the pulse profile shape changed from a double-peaked profile to one in which only the leading peak was detected. Figure 4 of \citet{levin10} shows that over the entire pulse profile at 1.4 and 3.1 GHz, when both peaks were detected, the polarized position angle (PPA) swung through 180 degrees. If an ATCA observation of PSR J1622--4950 takes place when the entire double-peaked pulse is switched on, and is therefore experiencing a very large PPA swing, phase-averaging of the pulse would produce a low linear polarization measurement, such as what we observed in 2008 Dec. However, if only part of the pulse is switched on, such as in the case of the last pulse profile shown in Figure 1 of \citet{levin10}, then a phased-averaged observation does not experience the full PPA swing. The phase-averaging would then result in less depolarization such as in the case of our 2008 Nov ATCA observation. 

The difference in circular polarization between the two ATCA epochs can be explained by the same changes in the pulse profile. In 2008 Dec, when we assume that the entire pulse profile is switched on, we see a significant fraction of circular polarization that was not observed in 2008 Nov. It is possible that the part of the pulse profile that was switched off during the 2008 Nov observation was the circularly polarized component, resulting in a lack of detectable circularly polarized emission during that epoch. This component then switched back on again in 2008 Dec, allowing us to detected circular polarization. Observations by \citet{levin10} have demonstrated that changes in the pulse profile of PSR J1622--4950 happen on the timescale of days and could therefore have occurred in the 14 days between the 2008 Nov and 2008 Dec ATCA observations, resulting in the observed polarization variability.

\subsection{Correlation between Time Variability in Radio and X-rays}

In order to determine if there is any correlation between the X-ray and radio emission from PSR J1622--4950, we have compared our X-ray and ATCA results with the Parkes 1.4 GHz light-curve reported by \citet{levin10}. Figure~\ref{fig3} shows the light-curve of PSR J1622--4950 where the radio 1.4 GHz flux values are in red (flux scale on the left axis) and X-ray flux values are in blue (flux scale on the right axis). The radio flux points include the Parkes measurements from \citet{levin10} and our ATCA detections extrapolated to 1.4 GHz. This extrapolation assumes that the radio spectral index of PSR J1622--4950, calculated from these ATCA observations (see Table~\ref{tab:2}), describes the radio spectrum down to low frequencies, just as in the case of XTE J1810--197 \citep{lazaridis08}. This results in estimated 1.4 GHz fluxes of $\sim39$ mJy during the 2008 Nov observation and $\sim69$ mJy during the 2008 Dec observation, which are far brighter than any other radio detection of PSR J1622--4950 at 1.4 GHz. If our values represent the fluxes of PSR J1622--4950 at 1.4 GHz on these two dates then this magnetar was undergoing a radio flaring event during our ATCA observations and was therefore in a high radio state. All other radio measurements, including those at earlier epochs than covered by the time range of Figure~\ref{fig3} \citep[see Figure 1 of][]{levin10}, have revealed considerably lower flux values or only upper limits.

The 2007 June (MJD 54264) \cxo\ observation detected PSR J1622--4950 in a reasonably high X-ray flux state when the magnetar was likely recovering from a recent X-ray outburst. The 2008 Nov and 2008 Dec ATCA observations then indicate a high radio flux from PSR J1622--4950 1.45 years after this \cxo\ observation (assuming that the spectral index from the ATCA observations can be extrapolated to 1.4 GHz). Studies of XTE J1810--197 and 1E J1547.0--5408 show that their radio emission was trigged by an X-ray outburst \citep{halpern05,camilo07v666} and, in the case of XTE J1810--197, this radio emission peaked in intensity $\sim3$ years after its X-ray outburst \citep{camilo06Nat}. Therefore, it is possible that the high radio state observed for PSR J1622--4950 with the ATCA in 2008 was triggered by the X-ray outburst that occurred around the time of the 2007 June \cxo\ observation.

The 2009 June and 2009 July \cxo\ observations show that the X-ray flux declined by a factor of 8 following the 2007 June \cxo\ observation. The 2009 June and 2009 July \cxo\ observations occurred during the period between MJD 54900 and MJD 55250 for which the Parkes observations showed the 1.4 GHz flux to be extremely variable (see Figure~\ref{fig3}). The \xmm\ observation shows that by 2011 Feb the X-ray flux had declined by an additional factor of 6. Further X-ray and radio observations are required to determine if PSR J1622--4950 has subsequently returned to a quiescent state.

An X-ray flux upper limit of $F_{x} \leq 3 \times 10^{-13}$ \erg\ ($0.3-10.0$ keV) was also obtained for PSR J1622--4950 in the 1999 April (MJD 51291) AGPS observation (see Section 3.1.1.). Figure 1 of \citet{levin10} indicates that Parkes did not detect PSR J1622--4950 at 1.4 GHz before, during or shortly after this \asca\ observation. It was not until 50 days after the \asca\ observation that Parkes detected radio emission. Unfortunately, without further X-ray flux history, it is not possible to determine if the X-ray detection is real and/or related to the subsequent radio detection.

\subsection{$\gamma$-ray Counterpart}

PSR J1622--4950 falls within the 95\% error circle of the Fermi Large Area Telescope (LAT) source \object[0FGL J1622.4-4945]{0FGL J1622.4--4945} \citep{abdo09}. 0FGL J1622.4--4945 was one of the Fermi-LAT bright $\gamma$-ray sources \citep{abdo09}, detected in the first three months of observations, with a statistical significance $\geq10 \sigma$. However, 0FGL J1622.4--4945 is one of 10 Fermi-LAT Bright Source List sources that were not detected in the Fermi-LAT First Source Catalog (1FGL), a more recent catalog in which each source detection is based on the average flux over an eleven month period with a statistical significance $\geq4 \sigma$ \citep{abdo10}. This is not unexpected due to its location in the Galactic ridge, since Fermi sources in this region are far more difficult to detect and characterize in the 1FGL catalog analysis \citep{abdo10}.

PSR J1622--4950 lies $\sim5.8'$ from the centroid of 0FGL J1622.4--4945 so it is worth investigating if there is an association between these two objects. Fermi $\gamma$-ray pulsars have been found to have spin-down luminosities between $\dot{E}=3 \times 10^{33}$ and $5 \times 10^{38}$ erg s$^{-1}$ \citep{abdo10psr}. While the spin-down luminosity of PSR J1622--4950 is encompassed in this range, this magnetar is far more distant than the $\gamma$-ray pulsars listed in \citet{abdo10psr}, the majority of which lie within 3kpc of Earth. The Fermi $\gamma$-ray pulsars have an $\dot{E}/d^{2}$ between $3 \times 10^{33}$ and $~1 \times 10^{38}$ erg s$^{-1}$ kpc$^{-2}$ \citep{abdo10psr}. The value of $\dot{E}/d^{2}$ for PSR J1622--4950 is $1 \times 10^{32}$ erg s$^{-1}$ kpc$^{-2}$, an order of magnitude smaller than the minimum value seen from the Fermi pulsars. An association between 0FGL J1622.4--4945 and PSR J1622--4950 is therefore unlikely, particularly as an investigation of much closer magnetars did not yield a $\gamma$-ray detection with Fermi-LAT \citep{abdo10mag}.

\subsection{Supernova Remnant Association}

Magnetars are young neutron stars. However, while the number of identified magnetars is increasing, few have been found to have a convincing association with a supernova remnant (SNR): \object[1E 2259+586]{1E 2259+586} in \object[SNR G109.1-01.0]{CTB 109} \citep{fahlman81}, \object[PSR J1841-0456]{1E 1841--045} in \object[SNR G027.3+00.0]{Kes 73} \citep{vasisht97}, \object[1E 1547.0-5408]{1E 1547.0--5408} in \object[SNR G327.2-00.1]{SNR G327.24--0.13} \citep{gelfand07} and the yet to be confirmed magnetars \object[AX J1845.0-0258]{AX J1845--0258} in \object[SNR G029.6+00.1]{SNR G29.6+0.1} \citep{gaensler99} and \object[CXOU J171405.7-381031]{CXOU J171405.7--381031} in \object[SNR G348.7+00.3]{CTB 37B} \citep{aharonian08,halpern10}.\footnote{Another possible association is \object[SGR 0526-66]{SGR 0526--66} with SNR N49 in the Large Magellanic Cloud \citep{cline82,gaensler01,park12}.} Such associations are important because SNRs provide independent constraints on the environment and the properties of the associated magnetar.

Source B (and its possibly associated counter-arc source C) appears to have similar properties to SNRs as the apparent lack of associated diffuse infrared emission, as mentioned in Section 3.2.1., implies a non-thermal nature \citep[for example see][]{brogan06}. PSR J1622--4950 also resides within $2.5'$ of the center of the extrapolated structure. Another pulsar, PSR J1622--4944, mentioned in Section 3.2.1., lies $\sim 1'$ from Source B but has a high characteristic age of $10^{6}$ yrs \citep{manchester01}. While it is true that a pulsar's characteristic age can be overestimated, it is extremely unusual for a pulsar with such a high characteristic age to be found associated with an SNR. In the absence of any further evidence, PSR J1622--4944 appears to be an ordinary old radio pulsar unrelated to source B. 

Confirmation of source B as an SNR requires the measurement of a non-thermal radio spectral index or the detection of linear polarization, neither of which is possible to obtain with our current data-sets. (SGPS lacks the sensitivity required to disentangle the diffuse sources in this complicated region and the MGPS does not have any polarization information. The frequencies of both these surveys are also too closely spaced to provide a meaningful spectral index estimation.) Instead we need to consider the feasibility of such an identification by exploring a possible connection between source B and PSR J1622--4950 using some of the pulsar/SNR association criteria established by \citet{kaspi96}. In the following discussion the criteria we explore are whether PSR J1622--4950 and G333.9+0.0 have consistent ages and if the implied transverse velocity of the magnetar away from the assumed explosion site is reasonable. As we do not know the distance to G333.9+0.0 we cannot investigate whether the distances to the magnetar and SNR are consistent. For the purposes of further consideration we assume a common distance of 9 kpc to both sources. The probability of chance alignment between the magnetar and SNR also needs to be considered as demonstrated by \citet{gaensler01}. We designate source B as G333.9+0.0 based on its approximate centroid and will assume it is an SNR in our discussion below.

The positional coincidence of PSR J1622--4950 with the center of G333.9+0.0 could suggest a possible association. The chance probability of finding an arc in MGPS1 whose emission is non-thermal (based on comparisons with GLIMPSE data), and whose center is within $2.5'$ of a given position on the sky, is about 5\% based upon inspecting 100 random positions for $315 \le l \le 357$ and $|b| \le 0.4$. This probability is not particularly high or low, and so does not strengthen or argue against an association.

The angular separation between the estimated center of the arc G333.9+0.0 and the position of the PSR J1622--4950 is $0.8-2.5'$. Using PSR J1622--4950's characteristic age of 4 kyr \citep{levin10}, this leads to a projected velocity of $500-1500$ km s$^{-1}$ for a distance of 9 kpc. This is consistent with the overall velocity distribution for pulsars \citep{arzoumanian02}, but higher than that observed or inferred for magnetars \citep{gaensler01,helfand07,kaplan09,deller12,tendulkar12}.

No X-ray emission is detected from G333.9+0.0 in the \xmm\ observation; the upper limit on the count-rate is 0.04 counts s$^{-1}$ using standard errors and roughly accounting for vignetting \citep{romer01}. Assuming a thin-shell morphology  and the Sedov-Taylor solution \citep{sedov46a,sedov46b,taylor50a,taylor50b}, we find that the upper limit implies a preshock ambient density lower than $n_{0} = 0.05$ cm$^{-3}$ for an explosion energy of $10^{51}$ erg. For a SNR with a 18 pc radius this ambient density predicts a swept up mass of 30 $M_{\odot}$, supporting Sedov-Taylor expansion. Here we have used the standard shock jump conditions for an ideal gas to determine the gas temperature based on the Sedov-Taylor age for the given value of $n_{0}$, and then estimated the X-ray count rate assuming a Raymond-Smith plasma model in \texttt{XSPEC}. We find a Sedov-Taylor upper-limit age of $6$ kyr, at most a factor of 1.5 greater than PSR J1622--4950's characteristic age of 4 kyr, making G333.9+0.0 potentially young and similar in age to convincing magnetar/SNR associations \citep{gaensler01}. This Sedov-Taylor age leads to an upper limit on shock velocity of $\sim 1200$ km s$^{-1}$. The associated proton temperature is $<2$ keV, indicating an electron temperature of $<0.2$ keV using the electron-ion equilibration relation from \citet{ghavamian07}. While the above upper limit on the density is much lower than the mean ISM value, it is not unreasonable for the low-density cavities formed from stellar winds of the massive stars that lead to core-collapse supernovae. We conclude that the radius of the shell and the absence of X-ray emission are both consistent with young SNRs in a low density medium.

The implied non-thermal nature of source A, combined with the positional coincidence of PSR J1622--4950 within its extent, raises the possibility that source A could be a pulsar wind nebula (PWN) generated by PSR J1622--4950. This would be an interesting result, as there is currently no known radio PWN associated with a magnetar. However, a radial profile analysis, comparing the detection of PSR J1622--4950 to the point-spread-function of the XMM image, do not show any evidence of an extended X-ray source. We also search for extended X-ray emission resulting from a dust-scattering halo, similar to that seen around 1E 1547.0--5408 \citep{tiengo10,olausen11}, but none was detected. ATCA observations, with longer integration times, are required to determine the true nature of source A.

The above do not provide direct evidence that G333.9+0.0 is the shell of an SNR associated with PSR J1622--4950, but there is no firm evidence to argue against such an association either. Another possible SNR association includes the ring of diffuse radio emission that forms part of source A, which sits $\sim2'$ south of PSR J1622--4950. \citet{levin10} discuss the possibility that this ring could be the parent SNR to PSR J1622--4950 as it appears non-thermal in nature given that it lacks an infrared counterpart. However, they consider a link unlikely given the ring's small size and the high implied magnetar birth velocity.

\section{Conclusions}

In this paper we have confirmed the \citet{levin10} magnetar identification of PSR J1622--4950 through the detection of significant X-ray flux variability and high X-ray luminosity. The high dynamic range in the X-ray flux, combined with the exponential characteristic decay time of $\tau = 360 \pm 11$ days, suggests that PSR J1622--4950 may be a new addition to the transient magnetar class, and could possibly be recovering from an X-ray outburst that occurred before or during the 2007 June \cxo\ observation. This X-ray flux variability, along with the variable radio flux and spectral index, make PSR J1622--4950 similar to the two other known radio magnetars, XTE J1810--197 and 1E 1547.0--5408. Observations with the Australia Telescope Compact Array show that PSR J1622--4950 may have undergone a radio flaring event approximately one and a half years after the 2007 June \cxo\ observation, which could have been triggered by the X-ray outburst that occurred around this time in 2007. The proximity of PSR J1622--4950 to the supernova remnant candidate G333.9+0.0, the implied transverse velocities for PSR J1622--4950, and the apparent young age of SNR G333.9+0.0, all support the possibility of a new magnetar/SNR association.



\asca\ observations in 1999 indicate that PSR J1622--4950 may be responsible for some of the X-ray flux detected from AX J162246--4956 in the \asca\ Galactic Plane Survey \citep{sugizaki01}. In the AGPS the magnetars XTE J1810--197, 1E 1547.0--5408, \object[1E 1841--045]{1E 1841--045} \citep{vasisht97}, and \object[SGR 1806-20]{SGR 1806--20} \citep{laros87,ulmer93}, were all detected as X-ray sources. Through our work in ChIcAGO we anticipate the discovery of other magnetars, which will allows us to further define the properties of this unusual population of neutron stars.

\begin{acknowledgements}

We thank the referee for their careful reading of the manuscript and constructive suggestions. G.E.A acknowledges the support of an Australian Postgraduate Award. B.M.G. acknowledges the support of an Australian Laureate Fellowship. P.O.S. acknowledges partial support from NASA Contract NAS8-03060. N.R. is supported by a Ramon y Cajal Research Fellowship to CSIC, and grants AYA2009-07391 and SGR2009-811, as well as the Formosa Program TW2010005. J.J.D was supported by NASA contract NAS8-39073 to the Chandra X-ray Center (CXC). P.E. acknowledges financial support from the Autonomous Region of Sardinia through a research grant under the program PO Sardegna FSE 2007--2013, L.R. 7/2007. D.T.H.S. acknowledges a STFC Advanced Fellowship. Support for this work was also provided by NASA through \cxo\ Award Number GO9-0155X issued by the CXC, which is operated by the Smithsonian Astrophysical Observatory for and on behalf of NASA. This research makes use of data obtained with the \cxo\ \textit{X-ray Observatory}, and software provided by the CXC in the application packages \texttt{CIAO}. Based on observations obtained with \textit{XMM-Newton}, an ESA science mission with instruments and contributions directly funded by ESA Member States and NASA. The MOST is operated with the support of the Australian Research Council and the Science Foundation for Physics within the University of Sydney. The ATCA and Parkes, part of the Australia Telescope, are funded by the Commonwealth of Australia for operation as a National Facility managed by CSIRO. Observing time on the 6.5m Baade Magellan Telescope, located at Las Campanas Observatory, was allocated through the Harvard-Smithsonian Center for Astrophysics. 2MASS is a joint project of the University of Massachusetts and the IPAC/Caltech, funded by the NASA and NFS. GLIMPSE survey data are part of the Spitzer Legacy Program. The \textit{Spitzer Space Telescope} is operated by the JPL/Caltech under a contract with NASA. This research has made use of NASA's Astrophysics Data System.

\end{acknowledgements}

{\it Facilities:} \facility{ASCA}, \facility{ATCA}, \facility{CXO (ACIS)}, \facility{Magellan:Baade (PANIC)}, \facility{Molonglo}, \facility{Parkes}, \facility{XMM (EPIC)}


\pagebreak

\input{table1.tex}

\input{table2.tex}

\input{table3.tex}

\input{figures.tex}

\end{document}

%% file: table1.tex
\begin{deluxetable}{lcccccccccccc}
\tablewidth{0pt}
\tabletypesize{\scriptsize}
\tablecaption{X-ray Observations of PSR J1622--4950\label{tab:1}}
\tablehead{
\colhead{Telescope} & \colhead{} & \colhead{Obs ID} & \colhead{} & \colhead{Date\tablenotemark{a}} & \colhead{} & \colhead{MJD\tablenotemark{c}} & \colhead{} & \colhead{Count-rate\tablenotemark{d}} & \colhead{} & \colhead{ACIS-S rate\tablenotemark{e}} & \colhead{} & \colhead{Off-axis Angle} 
\\
\cline{1-1} \cline{3-3} \cline{5-5} \cline{7-7} \cline{9-9} \cline{11-11} \cline{13-13}
\\
\colhead{} & \colhead{} & \colhead{Instrument} & \colhead{} & \colhead{Exp Time (ks)\tablenotemark{b}} & \colhead{} & \colhead{} & \colhead{} & \colhead{counts ks$^{-1}$} &  \colhead{} & \colhead{counts ks$^{-1}$} & \colhead{} & \colhead{arcmin}
}
\startdata
\cxo\ & & 8161 & & 2007-06-13 & & 54264 & & $96.8 \pm 6.9$ & & 96.8 & & 4.0
\\
& & ACIS-S & & 2.02 & & & & & & & &
\\
\\
\cxo\ & & 9911 & & 2009-06-14 & & 54996 & & $9.7 \pm 0.6$ & & 10.8 & & 16.1
\\
& & ACIS-I & & 60.10 & & & & & & & &
\\
\\
\cxo\ & & 10929 & & 2009-07-10 & & 55022 & & $7.1 \pm 0.6$ & & 9.3 & & 1.8
\\
& & ACIS-I & & 19.90 & & & & & & & &
\\
\\
\xmm\ & & 0654110101 & & 2011-02-22 & & 55615 & & $5.8 \pm 0.6$ & & 1.8 & & 1.1
\\
& & EPIC-PN & & 46.4 & & & & & & & &
\enddata

\tablenotetext{a}{The date is in the form yyyy-mm-dd.}

\tablenotetext{b}{The quoted exposure times are the effective exposure time after time intervals when there was flaring or when the source was on the chip gap have been removed.}

\tablenotetext{c}{Modified Julian Date}

\tablenotetext{d}{Total observed count-rate in the $0.3-8.0$ keV energy range.}

\tablenotetext{e}{Model predicted ACIS-S equivalent count-rate when compared to the 2007 June \cxo\ observation in the $0.3-8.0$ keV energy range.}

\end{deluxetable}

%% file: table2.tex
\begin{deluxetable}{ccccccccccccc}
\tablewidth{0pt}
\tabletypesize{\scriptsize}
\tablecaption{ATCA Observations of PSR J1622--4950\label{tab:2}}
\tablehead{
\colhead{Date} & \colhead{} & \colhead{MJD} & \colhead{} & \colhead{Central Frequency} & \colhead{} & \colhead{Flux Density} & \colhead{} & \colhead{Spectral Index} & \colhead{} & \multicolumn{3}{c}{Polarization}
\\
\cline{1-1} \cline{3-3} \cline{5-5} \cline{7-7} \cline{9-9} \cline{11-13}
\\
\colhead{(yyyy-mm-dd)} & \colhead{} & \colhead{} & \colhead{} & \colhead{(MHz)} & \colhead{} & \colhead{(mJy)} & \colhead{} & \colhead{($\alpha$)} & \colhead{} & \colhead{Linear (mJy)} & \colhead{PA (deg)} & \colhead{Circular (mJy)}
}
\startdata
2008-11-22 & & 54793 & & 5312 & & 33.0 $\pm$ 0.3 & & -0.13 $\pm$ 0.04 & & $26.6 \pm 0.7$ ($79\%$) & -17.5 $\pm 0.5$ & $\lesssim2.0$ ($\lesssim6$\%)
\\
 & & & & 8768 & & 30.9 $\pm$ 0.6 & & & & $25.0 \pm 0.8$ ($81\%$) & -25.8 $\pm 0.7$ & $\lesssim2.5$ ($\lesssim8$\%)
\\
\cline{1-13}
\\
2008-12-05 & & 54806 & & 4800 & & 40.4 $\pm$ 0.3 & & -0.44 $\pm$ 0.04 & & $5.7 \pm 0.4$ ($14\%$) & +26.7 $\pm 1.5$ & -$6.2 \pm$ 0.3 ($15\%$)
\\
 & & & & 8256 & & 31.9 $\pm$ 0.6 & & & & $5.8 \pm 0.7$ ($18\%$) & -22.5 $\pm 2.5$ & -$4.8 \pm$ 0.5 ($15\%$)
\\
\cline{1-13}
\\
2009-12-08 & & 55174 & & 5500 & & $13 \pm 1$ & & $+0.2 \pm 0.2$ & & & &
\\
2010-02-27\tablenotemark{a} & & 55255 & & 9000 & & $14.3 \pm 0.8$ & & & & & &
\enddata

\tablecomments{All errors are 1$\sigma$. The circular polarization upper bounds for the 2008-11-22 observations are limits on the magnitude.}

\tablenotetext{a}{The flux densities quoted are the average values at 5.5 and 9 GHz from the 2009-12-08 and 2010-02-27 ATCA observations, taken from \citet{levin10}.}


\end{deluxetable}

%% file: table3.tex
\begin{deluxetable}{lcccccccc}
\tablewidth{0pt}
\tabletypesize{\scriptsize}
\tablecaption{X-ray Spectral modeling of PSR J1622--4950\label{tab:3}}
\tablehead{
\colhead{Telescope} & \colhead{} & \multicolumn{3}{c}{Absorbed Blackbody Fit\tablenotemark{a}} & \colhead{} & \multicolumn{3}{c}{Absorbed Power-law Fit\tablenotemark{a}} 
\\
\cline{1-1} \cline{3-5} \cline{7-9}
\\
\colhead{Date} & \colhead{} & \colhead{$kT$} & \colhead{$F_{x,abs}$} & \colhead{$F_{x,unab}$}  & \colhead{} & \colhead{$\Gamma$} & \colhead{$F_{x,abs}$} & \colhead{$F_{x,unab}$}
\\
\colhead{} & \colhead{} & \colhead{$N_{H}$} & \colhead{$\chi^{2}_{red}$} & \colhead{$L_{x,unab}$}  & \colhead{} & \colhead{$N_{H}$} & \colhead{$\chi^{2}_{red}$} & \colhead{$L_{x,unab}$}
}
\startdata
\cxo\ & & $0.7 \pm 0.1$ & $1.4 \pm 0.3 \times 10^{-12}$ & $3.5^{+1.4}_{-0.9} \times 10^{-12}$ & & $4.2^{+1.0}_{-0.8}$ & $1.5^{+0.4}_{-0.3} \times 10^{-12}$ & $3.3^{+24.4}_{-2.7} \times 10^{-10}$
\\
2007-06-13 & & $5.4^{+1.6}_{-1.4}$ & 0.7 & $3.4^{+1.4}_{-0.9} \times 10^{34}$ & & $10.5^{+2.5}_{-2.1}$ & 0.7 & $3.2^{+23.9}_{-2.6} \times 10^{36}$
\\
\\
\cxo\ & & $0.7 \pm 0.1$ & $1.8^{+0.3}_{-0.2} \times 10^{-13}$ & $4.9^{+2.2}_{-1.2} \times 10^{-13}$ & & $4.3^{+0.9}_{-0.8}$ & $2.1^{+0.4}_{-0.3} \times 10^{-13}$ & $5.9^{+45.1}_{-4.9} \times 10^{-11}$
\\
2009-06-14 & & $5.4^{+1.6}_{-1.4}$ & & $4.8^{+2.2}_{-1.2} \times 10^{33}$ & & $10.5^{+2.5}_{-2.1}$ & & $5.7^{+43.8}_{-4.8} \times 10^{35}$
\\
\\
\cxo\ & & $0.8^{+0.3}_{-0.2}$ & $1.6^{+0.6}_{-0.4} \times 10^{-13}$ & $3.6^{+1.6}_{-0.9} \times 10^{-13}$ & & $3.6^{+1.2}_{-1.1}$ & $2.0^{+0.9}_{-0.6} \times 10^{-13}$ & $1.3^{+12.6}_{-1.1} \times 10^{-11}$
\\
2009-07-10 & & $5.4^{+1.6}_{-1.4}$ & & $3.5^{+1.6}_{-0.9} \times 10^{33}$ & & $10.5^{+2.5}_{-2.1}$ & & $1.2^{+12.2}_{-1.0} \times 10^{35}$
\\
\\
\xmm\ & & $0.5 \pm 0.1$ & $3.0^{+0.8}_{-0.6} \times 10^{-14}$ & $1.1^{+0.9}_{-0.4} \times 10^{-13}$ & & $5.4^{+1.3}_{-1.1}$ & $3.2^{+0.9}_{-0.8} \times 10^{-14}$ & $7.9^{+133.3}_{-7.1} \times 10^{-11}$
\\
2011-02-22 & & $5.4^{+1.6}_{-1.4}$ & & $1.1^{+0.9}_{-0.4} \times 10^{33}$ & & $10.5^{+2.5}_{-2.1}$ & & $7.7^{+129.4}_{-6.9} \times 10^{35}$
\enddata

\tablenotetext{a}{The best fit absorbed blackbody and absorbed power-law model parameters including temperature, $kT$ (keV), spectral index, $\Gamma$, absorption column density, $N_{H}$ ($10^{22}$ cm$^{-2}$),  and the reduced chi-square, $\chi_{red}^{2}$, from Chi Gehrels statistics. The absorbed and unabsorbed  X-ray flux, $F_{x,abs}$ and $F_{x,unab}$ (erg cm$^{-2}$ s$^{-1}$) respectively, as well as the unabsorbed X-ray luminosity, $L_{x,unab}$ (erg s$^{-1}$), are over the $0.3-10.0$ keV energy range. The unabsorbed luminosity was calculated assuming a distance of 9 kpc \citep{levin10}. All fit parameter errors are for 90\% confidence.}

\end{deluxetable}

%% file: figures.tex
\begin{figure}
\begin{center}
\includegraphics[width=\textwidth]{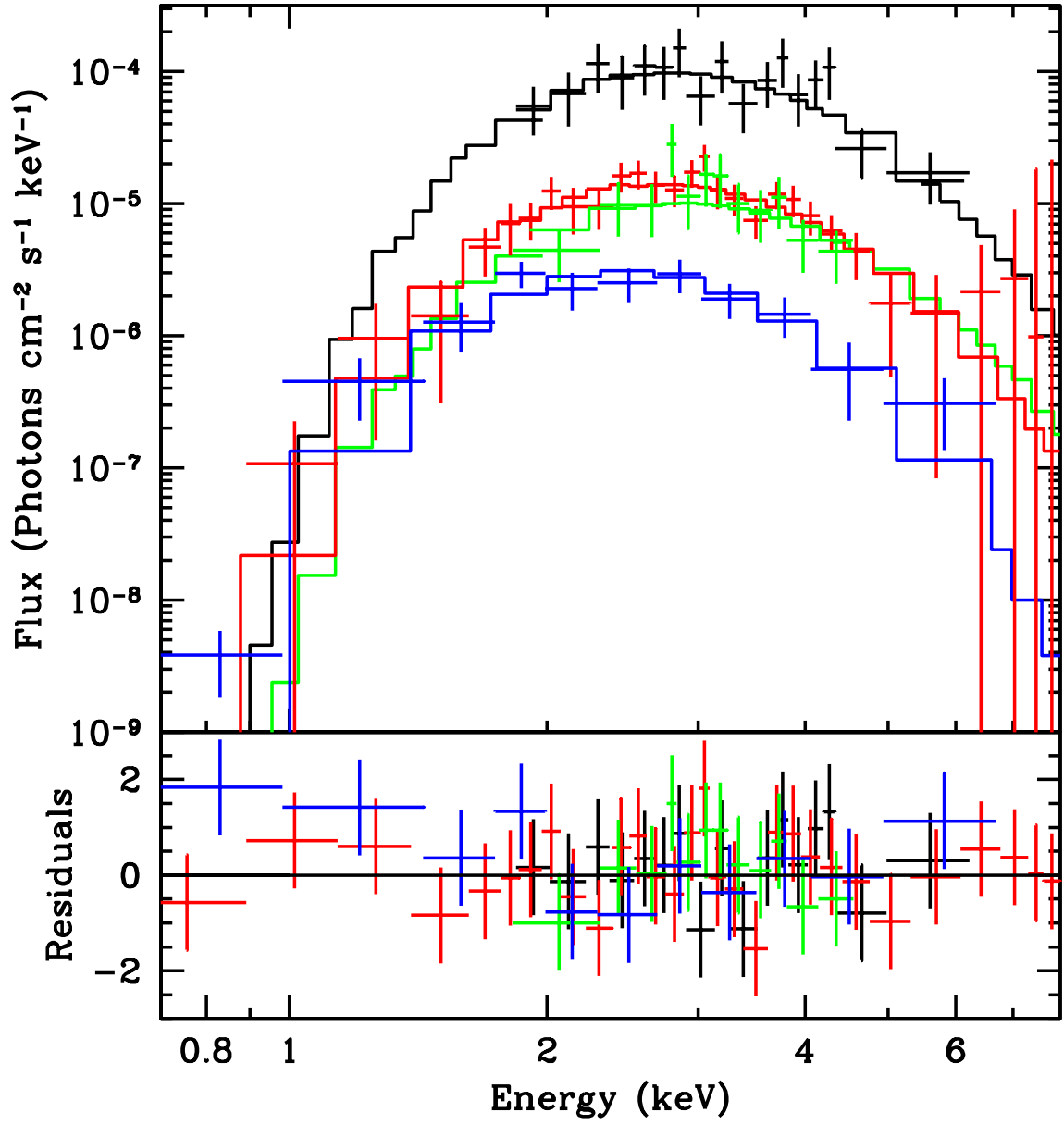}
\caption{The 2007 June (black), 2009 June (red) and 2009 July (green) \cxo\ spectra and 2011 Feb (blue) \xmm\ spectrum of PSR J1622--4950. These spectra were fit simultaneously with an absorbed black body model with $N_{H}$ locked between epochs but with $kT$ and the normalization values for each spectra allowed to vary individually. The bottom panel shows the residuals of these fits.}
\label{fig1}
\end{center}
\end{figure}

\begin{figure}
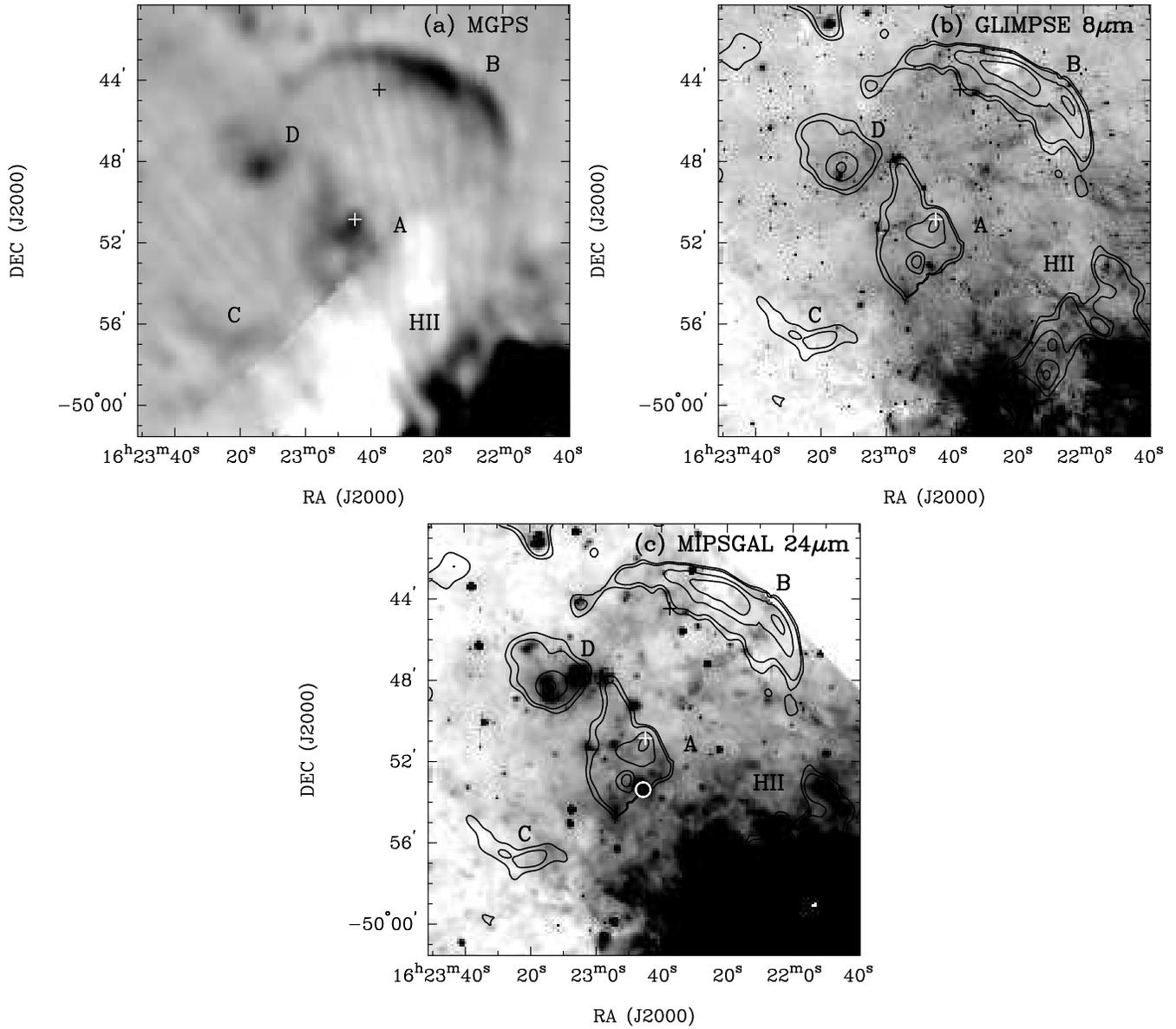

\begin{center}
\subfigure{\label{f2a}\includegraphics[width=0.43\textwidth, angle=270]{f2a.eps}}
\subfigure{\label{f2b}\includegraphics[width=0.43\textwidth, angle=270]{f2b.eps}}
\subfigure{\label{f2c}\includegraphics[width=0.43\textwidth, angle=270]{f2c.eps}}
\caption{Radio and infrared images of the region surrounding PSR J1622--4950. Each image is centered on the position of PSR J1622--4950, indicated by the white ``+'' symbol. Nearby radio sources are labeled A, B, C, and D, and the H{\sc ii} region, G333.6--0.2, is also indicated. The black contours in panels b and c show the MGPS1 0.005, 0.01, 0.03 and 0.05 mJy per beam radio emission levels. The nearby pulsar, PSR J1622--4944, is indicated by a black ``+'' symbol. a) The surrounding region as seen in the MGPS1 radio survey at 843 MHz and at a resolution of $43''$. b) The grayscale is the surrounding region as seen in GLIMPSE at 8 {\micron} at a resolution of $1.2''$. c) The grayscale as seen by MIPSGAL at 24 {\micron} at a resolution of $6''$. The white circle indicates the position of IRAS 16190--4946 with a radius of $23''$, equivalent to its major axis position uncertainty.}
\label{fig2}
\end{center}
\end{figure}

\pagebreak

\begin{figure}
\begin{center}
\includegraphics[width=\textwidth]{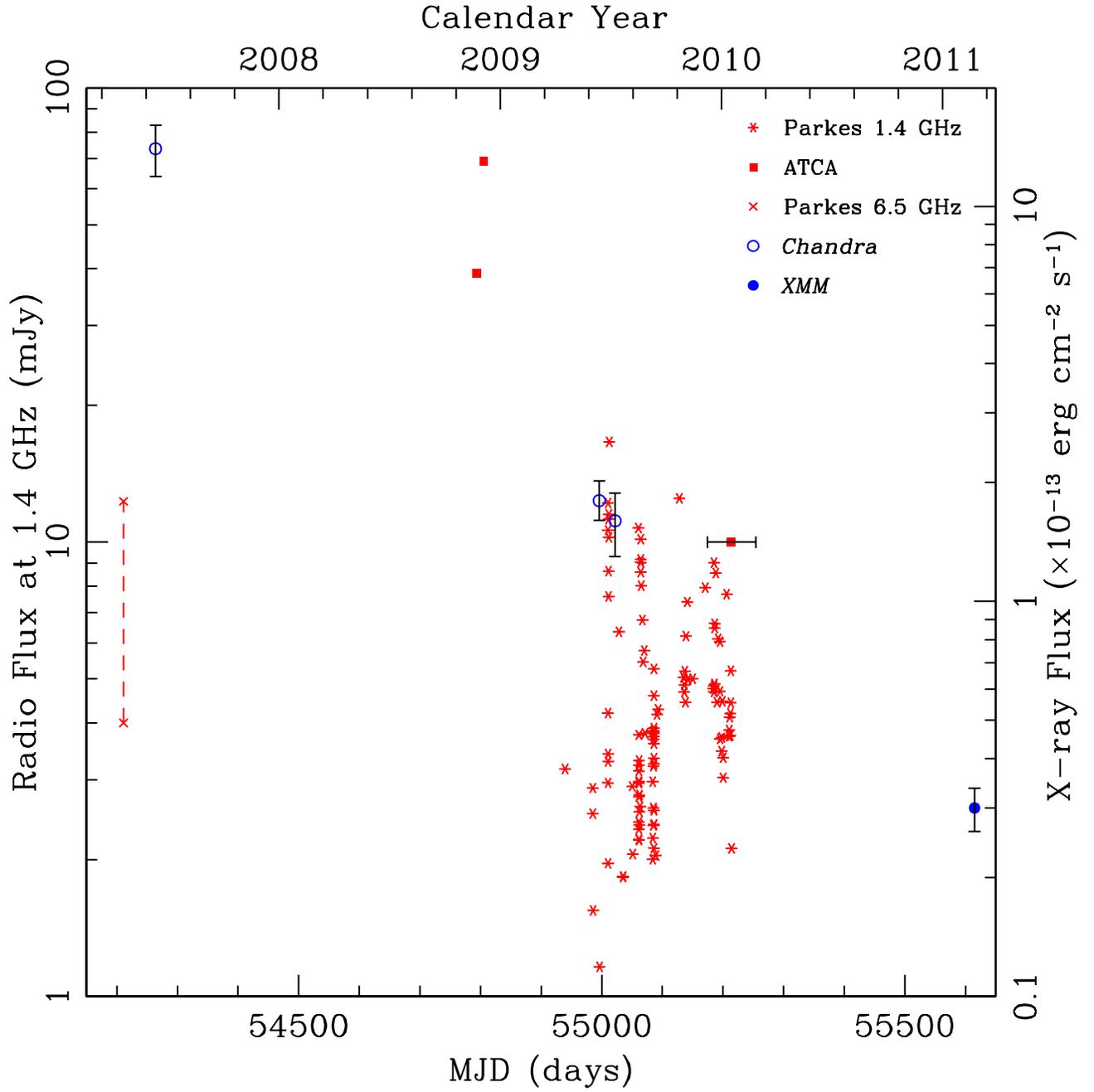}
\caption{Radio and X-ray light-curves of PSR J1622--4950 over 1500 days. The left vertical axis shows the radio flux at 1.4 GHz in units of mJy and the right vertical axis shows the absorbed X-ray flux in units of  $10^{-13}$ erg cm$^{-2}$ s$^{-1}$ in the $0.3-10.0$ keV energy range. The radio flux points are in red where the Parkes 1.4 GHz data are denoted by asterisks and the Parkes 6.5 GHz Multibeam pulsar survey measurement is a range of possible values at 1.4 GHz, calculated from a range of possible radio spectral indices, denoted by two crosses connected by a dashed line. \citep[These light-curve points were originally depicted in Figure 1 of][]{levin10}. The red squares show the ATCA detections of PSR J1622-4950 when extrapolated to 1.4 GHz. The ATCA detection with the horizontal error bar is the average flux value from two ATCA observations, taken in 2009 Dec and 2010 Feb, published by \citet{levin10}. The ATCA flux errors are the size of the data point. The X-ray flux points are in blue where the \cxo\ observations are denoted by open circles and the \xmm\ observation by a filled circle. The error bars are $1\sigma$. The upper-limit on the X-ray flux of PSR J1622--4950 detected by \asca\ in 1999 April (MJD 51291) was $F_{x} \leq 3 \times 10^{-13}$ \erg.}
\label{fig3}
\end{center}
\end{figure}

%% file: ms.bbl
\begin{thebibliography}{98}
\expandafter\ifx\csname natexlab\endcsname\relax\def\natexlab#1{#1}\fi

\bibitem[{{Abdo} {et~al.}(2010{\natexlab{a}}){Abdo}, {Ackermann}, {Ajello},
  {Allafort}, {Antolini}, {Atwood}, {Axelsson}, {Baldini}, {Ballet},
  {Barbiellini}, {Bastieri}, {Baughman}, {Bechtol}, {Bellazzini}, {Belli},
  {Berenji}, {Bisello}, {Blandford}, {Bloom}, {Bonamente}, {Bonnell},
  {Borgland}, {Bouvier}, {Bregeon}, {Brez}, {Brigida}, {Bruel}, {Burnett},
  {Busetto}, {Buson}, {Caliandro}, {Cameron}, {Campana}, {Canadas}, {Caraveo},
  {Carrigan}, {Casandjian}, {Cavazzuti}, {Ceccanti}, {Cecchi}, {{\c C}elik},
  {Charles}, {Chekhtman}, {Cheung}, {Chiang}, {Cillis}, {Ciprini}, {Claus},
  {Cohen-Tanugi}, {Conrad}, {Corbet}, {Davis}, {DeKlotz}, {den Hartog},
  {Dermer}, {de Angelis}, {de Luca}, {de Palma}, {Digel}, {Dormody}, {Silva},
  {Drell}, {Dubois}, {Dumora}, {Fabiani}, {Farnier}, {Favuzzi}, {Fegan},
  {Ferrara}, {Focke}, {Fortin}, {Frailis}, {Fukazawa}, {Funk}, {Fusco},
  {Gargano}, {Gasparrini}, {Gehrels}, {Germani}, {Giavitto}, {Giebels},
  {Giglietto}, {Giommi}, {Giordano}, {Giroletti}, {Glanzman}, {Godfrey},
  {Grenier}, {Grondin}, {Grove}, {Guillemot}, {Guiriec}, {Gustafsson},
  {Hadasch}, {Hanabata}, {Harding}, {Hayashida}, {Hays}, {Healey}, {Hill},
  {Horan}, {Hughes}, {Iafrate}, {J{\'o}hannesson}, {Johnson}, {Johnson},
  {Johnson}, {Johnson}, {Kamae}, {Katagiri}, {Kataoka}, {Kawai}, {Kerr},
  {Kn{\"o}dlseder}, {Kocevski}, {Kuss}, {Lande}, {Landriu}, {Latronico}, {Lee},
  {Lemoine-Goumard}, {Lionetto}, {Llena Garde}, {Longo}, {Loparco}, {Lott},
  {Lovellette}, {Lubrano}, {Madejski}, {Makeev}, {Marangelli}, {Marelli},
  {Massaro}, {Mazziotta}, {McConville}, {McEnery}, {Michelson}, {Minuti},
  {Mitthumsiri}, {Mizuno}, {Moiseev}, {Mongelli}, {Monte}, {Monzani},
  {Moretti}, {Morselli}, {Moskalenko}, {Murgia}, {Nakajima}, {Nakamori},
  {Naumann-Godo}, {Nolan}, {Norris}, {Nuss}, {Ohno}, {Ohsugi}, {Omodei},
  {Orlando}, {Ormes}, {Ozaki}, {Paccagnella}, {Paneque}, {Panetta}, {Parent},
  {Pelassa}, {Pepe}, {Pesce-Rollins}, {Pinchera}, {Piron}, {Porter}, {Poupard},
  {Rain{\`o}}, {Rando}, {Ray}, {Razzano}, {Razzaque}, {Rea}, {Reimer},
  {Reimer}, {Reposeur}, {Ripken}, {Ritz}, {Rochester}, {Rodriguez}, {Romani},
  {Roth}, {Sadrozinski}, {Salvetti}, {Sanchez}, {Sander}, {Saz Parkinson},
  {Scargle}, {Schalk}, {Scolieri}, {Sgr{\`o}}, {Shaw}, {Siskind}, {Smith},
  {Smith}, {Spandre}, {Spinelli}, {Starck}, {Stephens}, {Striani}, {Strickman},
  {Strong}, {Suson}, {Tajima}, {Takahashi}, {Takahashi}, {Tanaka}, {Thayer},
  {Thayer}, {Thompson}, {Tibaldo}, {Tibolla}, {Tinebra}, {Torres}, {Tosti},
  {Tramacere}, {Uchiyama}, {Usher}, {Van Etten}, {Vasileiou}, {Vilchez},
  {Vitale}, {Waite}, {Wallace}, {Wang}, {Watters}, {Winer}, {Wood}, {Yang},
  {Ylinen}, \& {Ziegler}}]{abdo10}
{Abdo}, A.~A. \etal . 2010{\natexlab{a}}, \apjs, 188, 405

\bibitem[{{Abdo} {et~al.}(2010{\natexlab{b}}){Abdo}, {Ackermann}, {Ajello},
  {Allafort}, {Baldini}, {Ballet}, {Barbiellini}, {Baring}, {Bastieri},
  {Bellazzini}, {Blandford}, {Bloom}, {Bonamente}, {Borgland}, {Bouvier},
  {Bregeon}, {Brigida}, {Bruel}, {Burnett}, {Caliandro}, {Cameron}, {Caraveo},
  {Cecchi}, {{\c C}elik}, {Chaty}, {Chekhtman}, {Cheung}, {Chiang}, {Ciprini},
  {Claus}, {Conrad}, {den Hartog}, {Dermer}, {de Angelis}, {de Palma}, {Dib},
  {Dormody}, {Silva}, {Drell}, {Dubois}, {Dumora}, {Enoto}, {Favuzzi},
  {Frailis}, {Fusco}, {Gargano}, {Gehrels}, {Giglietto}, {Giommi}, {Giordano},
  {Giroletti}, {Glanzman}, {Godfrey}, {Grenier}, {Grondin}, {Guiriec},
  {Hadasch}, {Hanabata}, {Harding}, {Hays}, {Israel}, {J{\'o}hannesson},
  {Johnson}, {Kaspi}, {Katagiri}, {Kataoka}, {Kn{\"o}dlseder}, {Kuss}, {Lande},
  {Lee}, {Lemoine-Goumard}, {Longo}, {Loparco}, {Lovellette}, {Lubrano},
  {Makeev}, {Marelli}, {Mazziotta}, {McEnery}, {Mehault}, {Michelson},
  {Mizuno}, {Moiseev}, {Monte}, {Monzani}, {Morselli}, {Moskalenko}, {Murgia},
  {Naumann-Godo}, {Nolan}, {Nuss}, {Ohsugi}, {Okumura}, {Omodei}, {Orlando},
  {Ormes}, {Ozaki}, {Paneque}, {Parent}, {Pepe}, {Pesce-Rollins}, {Piron},
  {Porter}, {Rain{\`o}}, {Rando}, {Razzano}, {Rea}, {Reimer}, {Reimer},
  {Reposeur}, {Ritz}, {Sadrozinski}, {Saz Parkinson}, {Sgr{\`o}}, {Siskind},
  {Smith}, {Spandre}, {Spinelli}, {Strickman}, {Takahashi}, {Tanaka}, {Thayer},
  {Thompson}, {Tibaldo}, {Torres}, {Tosti}, {Tramacere}, {Troja}, {Uchiyama},
  {Usher}, {Vandenbroucke}, {Vasileiou}, {Vianello}, {Vitale}, {Waite},
  {Winer}, {Wood}, {Yang}, \& {Ziegler}}]{abdo10mag}
---. 2010{\natexlab{b}}, \apjl, 725, L73

\bibitem[{{Abdo} {et~al.}(2009){Abdo}, {Ackermann}, {Ajello}, {Atwood},
  {Axelsson}, {Baldini}, {Ballet}, {Band}, {Barbiellini}, {Bastieri},
  {Battelino}, {Baughman}, {Bechtol}, {Bellazzini}, {Berenji}, {Bignami},
  {Blandford}, {Bloom}, {Bonamente}, {Borgland}, {Bouvier}, {Bregeon}, {Brez},
  {Brigida}, {Bruel}, {Burnett}, {Caliandro}, {Cameron}, {Caraveo},
  {Casandjian}, {Cavazzuti}, {Cecchi}, {Charles}, {Chekhtman}, {Cheung},
  {Chiang}, {Ciprini}, {Claus}, {Cohen-Tanugi}, {Cominsky}, {Conrad}, {Corbet},
  {Costamante}, {Cutini}, {Davis}, {Dermer}, {de Angelis}, {de Luca}, {de
  Palma}, {Digel}, {Dormody}, {do Couto e Silva}, {Drell}, {Dubois}, {Dumora},
  {Farnier}, {Favuzzi}, {Fegan}, {Ferrara}, {Focke}, {Frailis}, {Fukazawa},
  {Funk}, {Fusco}, {Gargano}, {Gasparrini}, {Gehrels}, {Germani}, {Giebels},
  {Giglietto}, {Giommi}, {Giordano}, {Glanzman}, {Godfrey}, {Grenier},
  {Grondin}, {Grove}, {Guillemot}, {Guiriec}, {Hanabata}, {Harding}, {Hartman},
  {Hayashida}, {Hays}, {Healey}, {Horan}, {Hughes}, {J{\'o}hannesson},
  {Johnson}, {Johnson}, {Johnson}, {Johnson}, {Kamae}, {Katagiri}, {Kataoka},
  {Kawai}, {Kerr}, {Kn{\"o}dlseder}, {Kocevski}, {Kocian}, {Komin}, {Kuehn},
  {Kuss}, {Lande}, {Latronico}, {Lee}, {Lemoine-Goumard}, {Longo}, {Loparco},
  {Lott}, {Lovellette}, {Lubrano}, {Madejski}, {Makeev}, {Marelli},
  {Mazziotta}, {McConville}, {McEnery}, {McGlynn}, {Meurer}, {Michelson},
  {Mitthumsiri}, {Mizuno}, {Moiseev}, {Monte}, {Monzani}, {Moretti},
  {Morselli}, {Moskalenko}, {Murgia}, {Nakamori}, {Nolan}, {Norris}, {Nuss},
  {Ohno}, {Ohsugi}, {Omodei}, {Orlando}, {Ormes}, {Ozaki}, {Paneque},
  {Panetta}, {Parent}, {Pelassa}, {Pepe}, {Pesce-Rollins}, {Piron}, {Porter},
  {Poupard}, {Rain{\`o}}, {Rando}, {Ray}, {Razzano}, {Rea}, {Reimer}, {Reimer},
  {Reposeur}, {Ritz}, {Rochester}, {Rodriguez}, {Romani}, {Roth}, {Ryde},
  {Sadrozinski}, {Sanchez}, {Sander}, {Saz Parkinson}, {Scargle}, {Schalk},
  {Sellerholm}, {Sgr{\`o}}, {Shaw}, {Shrader}, {Sierpowska-Bartosik},
  {Siskind}, {Smith}, {Smith}, {Spandre}, {Spinelli}, {Starck}, {Stephens},
  {Strickman}, {Strong}, {Suson}, {Tajima}, {Takahashi}, {Takahashi}, {Tanaka},
  {Thayer}, {Thayer}, {Thompson}, {Tibaldo}, {Tibolla}, {Torres}, {Tosti},
  {Tramacere}, {Uchiyama}, {Usher}, {Van Etten}, {Vilchez}, {Vitale}, {Waite},
  {Wallace}, {Wang}, {Watters}, {Winer}, {Wood}, {Ylinen}, {Ziegler}, \& {The
  Fermi/LAT Collaboration}}]{abdo09}
---. 2009, \apjs, 183, 46

\bibitem[{{Abdo} {et~al.}(2010{\natexlab{c}}){Abdo}, {Ackermann}, {Ajello},
  {Atwood}, {Axelsson}, {Baldini}, {Ballet}, {Barbiellini}, {Baring},
  {Bastieri}, \& et~al.}]{abdo10psr}
---. 2010{\natexlab{c}}, \apjs, 187, 460

\bibitem[{{Aharonian} {et~al.}(2008){Aharonian}, {Akhperjanian}, {Barres de
  Almeida}, {Bazer-Bachi}, {Behera}, {Beilicke}, {Benbow}, {Bernl{\"o}hr},
  {Boisson}, {Borrel}, {Braun}, {Brion}, {Brucker}, {B{\"u}hler}, {Bulik},
  {B{\"u}sching}, {Boutelier}, {Carrigan}, {Chadwick}, {Chaves}, {Chounet},
  {Clapson}, {Coignet}, {Cornils}, {Costamante}, {Dalton}, {Degrange},
  {Dickinson}, {Djannati-Ata{\"i}}, {Domainko}, {O'C.~Drury}, {Dubois},
  {Dubus}, {Dyks}, {Egberts}, {Emmanoulopoulos}, {Espigat}, {Farnier},
  {Feinstein}, {Fiasson}, {F{\"o}rster}, {Fontaine}, {Funk}, {F{\"u}{\ss}ling},
  {Gabici}, {Gallant}, {Giebels}, {Glicenstein}, {Gl{\"u}ck}, {Goret},
  {Hadjichristidis}, {Hauser}, {Hauser}, {Heinzelmann}, {Henri}, {Hermann},
  {Hinton}, {Hoffmann}, {Hofmann}, {Holleran}, {Hoppe}, {Horns},
  {Jacholkowska}, {de Jager}, {Jung}, {Katarzy{\'n}ski}, {Kaufmann},
  {Kendziorra}, {Kerschhaggl}, {Khangulyan}, {Kh{\'e}lifi}, {Keogh}, {Komin},
  {Kosack}, {Lamanna}, {Latham}, {Lemoine-Goumard}, {Lenain}, {Lohse},
  {Martin}, {Martineau-Huynh}, {Marcowith}, {Masterson}, {Maurin}, {McComb},
  {Moderski}, {Moulin}, {Naumann-Godo}, {de Naurois}, {Nedbal}, {Nekrassov},
  {Nolan}, {Ohm}, {Olive}, {de O{\~n}a Wilhelmi}, {Orford}, {Osborne},
  {Ostrowski}, {Panter}, {Pedaletti}, {Pelletier}, {Petrucci}, {Pita},
  {P{\"u}hlhofer}, {Punch}, {Quirrenbach}, {Raubenheimer}, {Raue}, {Rayner},
  {Renaud}, {Rieger}, {Reimer}, {Ripken}, {Rob}, {Rosier-Lees}, {Rowell},
  {Rudak}, {Ruppel}, {Sahakian}, {Santangelo}, {Schlickeiser}, {Sch{\"o}ck},
  {Schr{\"o}der}, {Schwanke}, {Schwarzburg}, {Schwemmer}, {Shalchi}, {Skilton},
  {Sol}, {Spangler}, {Stawarz}, {Steenkamp}, {Stegmann}, {Superina}, {Tam},
  {Tavernet}, {Terrier}, {van Eldik}, {Vasileiadis}, {Venter}, {Vialle},
  {Vincent}, {Vivier}, {V{\"o}lk}, {Volpe}, {Wagner}, {Ward}, {Zdziarski}, \&
  {Zech}}]{aharonian08}
{Aharonian}, F. \etal . 2008, \aap, 486, 829

\bibitem[{{Anderson} {et~al.}(2011){Anderson}, {Gaensler}, {Kaplan}, {Posselt},
  {Slane}, {Murray}, {Mauerhan}, {Benjamin}, {Brogan}, {Chakrabarty}, {Drake},
  {Drew}, {Grindlay}, {Hong}, {Lazio}, {Lee}, {Steeghs}, \& {van
  Kerkwijk}}]{anderson11}
{Anderson}, G.~E. \etal . 2011, \apj, 727, 105

\bibitem[{{Arabadjis} \& {Bregman}(1999)}]{arabadjis99}
{Arabadjis}, J.~S. \& {Bregman}, J.~N. 1999, \apj, 510, 806

\bibitem[{{Arzoumanian} {et~al.}(2002){Arzoumanian}, {Chernoff}, \&
  {Cordes}}]{arzoumanian02}
{Arzoumanian}, Z., {Chernoff}, D.~F., \& {Cordes}, J.~M. 2002, \apj, 568, 289

\bibitem[{{Balucinska-Church} \& {McCammon}(1992)}]{balucinska92}
{Balucinska-Church}, M. \& {McCammon}, D. 1992, \apj, 400, 699

\bibitem[{{Beloborodov} \& {Thompson}(2007)}]{beloborodov07}
{Beloborodov}, A.~M. \& {Thompson}, C. 2007, \apj, 657, 967

\bibitem[{{Benjamin} {et~al.}(2003){Benjamin}, {Churchwell}, {Babler}, {Bania},
  {Clemens}, {Cohen}, {Dickey}, {Indebetouw}, {Jackson}, {Kobulnicky},
  {Lazarian}, {Marston}, {Mathis}, {Meade}, {Seager}, {Stolovy}, {Watson},
  {Whitney}, {Wolff}, \& {Wolfire}}]{benjamin03}
{Benjamin}, R.~A. \etal . 2003, \pasp, 115, 953

\bibitem[{{Bernardini} {et~al.}(2009){Bernardini}, {Israel}, {Dall'Osso},
  {Stella}, {Rea}, {Zane}, {Turolla}, {Perna}, {Falanga}, {Campana},
  {G{\"o}tz}, {Mereghetti}, \& {Tiengo}}]{bernardini09}
{Bernardini}, F. \etal . 2009, \aap, 498, 195

\bibitem[{{Bernardini} {et~al.}(2011){Bernardini}, {Israel}, {Stella},
  {Turolla}, {Esposito}, {Rea}, {Zane}, {Tiengo}, {Campana}, {G{\"o}tz},
  {Mereghetti}, \& {Romano}}]{bernardini11}
---. 2011, \aap, 529, A19+

\bibitem[{{Bertin} \& {Arnouts}(1996)}]{bertain96}
{Bertin}, E. \& {Arnouts}, S. 1996, \aaps, 117, 393

\bibitem[{{Brogan} {et~al.}(2006){Brogan}, {Gelfand}, {Gaensler}, {Kassim}, \&
  {Lazio}}]{brogan06}
{Brogan}, C.~L., {Gelfand}, J.~D., {Gaensler}, B.~M., {Kassim}, N.~E., \&
  {Lazio}, T.~J.~W. 2006, \apjl, 639, L25

\bibitem[{{Camilo} {et~al.}(2007{\natexlab{a}}){Camilo}, {Cognard}, {Ransom},
  {Halpern}, {Reynolds}, {Zimmerman}, {Gotthelf}, {Helfand}, {Demorest},
  {Theureau}, \& {Backer}}]{camilo07v663}
{Camilo}, F. \etal . 2007{\natexlab{a}}, \apj, 663, 497

\bibitem[{{Camilo} {et~al.}(2007{\natexlab{b}}){Camilo}, {Ransom}, {Halpern},
  \& {Reynolds}}]{camilo07v666}
{Camilo}, F., {Ransom}, S.~M., {Halpern}, J.~P., \& {Reynolds}, J.
  2007{\natexlab{b}}, \apjl, 666, L93

\bibitem[{{Camilo} {et~al.}(2006){Camilo}, {Ransom}, {Halpern}, {Reynolds},
  {Helfand}, {Zimmerman}, \& {Sarkissian}}]{camilo06Nat}
{Camilo}, F., {Ransom}, S.~M., {Halpern}, J.~P., {Reynolds}, J., {Helfand},
  D.~J., {Zimmerman}, N., \& {Sarkissian}, J. 2006, \nat, 442, 892

\bibitem[{{Camilo} {et~al.}(2007{\natexlab{c}}){Camilo}, {Ransom},
  {Pe{\~n}alver}, {Karastergiou}, {van Kerkwijk}, {Durant}, {Halpern},
  {Reynolds}, {Thum}, {Helfand}, {Zimmerman}, \& {Cognard}}]{camilo07v669}
{Camilo}, F. \etal . 2007{\natexlab{c}}, \apj, 669, 561

\bibitem[{{Camilo} {et~al.}(2008){Camilo}, {Reynolds}, {Johnston}, {Halpern},
  \& {Ransom}}]{camilo08}
{Camilo}, F., {Reynolds}, J., {Johnston}, S., {Halpern}, J.~P., \& {Ransom},
  S.~M. 2008, \apj, 679, 681

\bibitem[{{Camilo} {et~al.}(2007{\natexlab{d}}){Camilo}, {Reynolds},
  {Johnston}, {Halpern}, {Ransom}, \& {van Straten}}]{camilo07v659}
{Camilo}, F., {Reynolds}, J., {Johnston}, S., {Halpern}, J.~P., {Ransom},
  S.~M., \& {van Straten}, W. 2007{\natexlab{d}}, \apjl, 659, L37

\bibitem[{{Carey} {et~al.}(2009){Carey}, {Noriega-Crespo}, {Mizuno}, {Shenoy},
  {Paladini}, {Kraemer}, {Price}, {Flagey}, {Ryan}, {Ingalls}, {Kuchar},
  {Pinheiro Gon{\c c}alves}, {Indebetouw}, {Billot}, {Marleau}, {Padgett},
  {Rebull}, {Bressert}, {Ali}, {Molinari}, {Martin}, {Berriman}, {Boulanger},
  {Latter}, {Miville-Deschenes}, {Shipman}, \& {Testi}}]{carey09}
{Carey}, S.~J. \etal . 2009, \pasp, 121, 76

\bibitem[{{Chatterjee} {et~al.}(2009){Chatterjee}, {Brisken}, {Vlemmings},
  {Goss}, {Lazio}, {Cordes}, {Thorsett}, {Fomalont}, {Lyne}, \&
  {Kramer}}]{chatterjee09}
{Chatterjee}, S. \etal . 2009, \apj, 698, 250

\bibitem[{{Cline} {et~al.}(1982){Cline}, {Desai}, {Teegarden}, {Evans},
  {Klebesadel}, {Laros}, {Barat}, {Hurley}, {Niel}, \& {Weisskopf}}]{cline82}
{Cline}, T.~L. \etal . 1982, \apjl, 255, L45

\bibitem[{{Cordes} \& {Lazio}(2002)}]{cordes02}
{Cordes}, J.~M. \& {Lazio}, T.~J.~W. 2002, arXiv:astro-ph/0207156

\bibitem[{{Deller} {et~al.}(2012){Deller}, {Camilo}, {Reynolds}, \&
  {Halpern}}]{deller12}
{Deller}, A.~T., {Camilo}, F., {Reynolds}, J.~E., \& {Halpern}, J.~P. 2012,
  arXiv:astro-ph/1201.4684

\bibitem[{{Deller} {et~al.}(2009){Deller}, {Tingay}, {Bailes}, \&
  {Reynolds}}]{deller09}
{Deller}, A.~T., {Tingay}, S.~J., {Bailes}, M., \& {Reynolds}, J.~E. 2009,
  \apj, 701, 1243

\bibitem[{{Dickey} \& {Lockman}(1990)}]{dickey90}
{Dickey}, J.~M. \& {Lockman}, F.~J. 1990, \araa, 28, 215

\bibitem[{{Dorman} \& {Arnaud}(2001)}]{dorman01}
{Dorman}, B. \& {Arnaud}, K.~A. 2001, in Astronomical Data Analysis Software
  and Systems X, ed. F.~R. {Harnden}, F.~A. {Primini~Jr}, \& H.~E. {Payne} (San
  Francisco: Astronomical Society of the Pacific), p. 415

\bibitem[{{Durant} \& {van Kerkwijk}(2005)}]{durant05}
{Durant}, M. \& {van Kerkwijk}, M.~H. 2005, \apj, 627, 376

\bibitem[{{Enoto} {et~al.}(2009){Enoto}, {Nakagawa}, {Rea}, {Esposito},
  {G{\"o}tz}, {Hurley}, {Israel}, {Kokubun}, {Makishima}, {Mereghetti},
  {Murakami}, {Nakazawa}, {Sakamoto}, {Stella}, {Tiengo}, {Turolla}, {Yamada},
  {Yamaoka}, {Yoshida}, \& {Zane}}]{enoto09}
{Enoto}, T. \etal . 2009, \apjl, 693, L122

\bibitem[{{Evans} {et~al.}(2010){Evans}, {Primini}, {Glotfelty}, {Anderson},
  {Bonaventura}, {Chen}, {Davis}, {Doe}, {Evans}, {Fabbiano}, {Galle}, {Gibbs},
  {Grier}, {Hain}, {Hall}, {Harbo}, {(Helen He}, {Houck}, {Karovska},
  {Kashyap}, {Lauer}, {McCollough}, {McDowell}, {Miller}, {Mitschang},
  {Morgan}, {Mossman}, {Nichols}, {Nowak}, {Plummer}, {Refsdal}, {Rots},
  {Siemiginowska}, {Sundheim}, {Tibbetts}, {Van Stone}, {Winkelman}, \&
  {Zografou}}]{evans10}
{Evans}, I.~N. \etal . 2010, \apjs, 189, 37

\bibitem[{{Fahlman} \& {Gregory}(1981)}]{fahlman81}
{Fahlman}, G.~G. \& {Gregory}, P.~C. 1981, \nat, 293, 202

\bibitem[{{Gaensler} {et~al.}(1999){Gaensler}, {Gotthelf}, \&
  {Vasisht}}]{gaensler99}
{Gaensler}, B.~M., {Gotthelf}, E.~V., \& {Vasisht}, G. 1999, \apjl, 526, L37

\bibitem[{{Gaensler} {et~al.}(2001){Gaensler}, {Slane}, {Gotthelf}, \&
  {Vasisht}}]{gaensler01}
{Gaensler}, B.~M., {Slane}, P.~O., {Gotthelf}, E.~V., \& {Vasisht}, G. 2001,
  \apj, 559, 963

\bibitem[{{Garmire} {et~al.}(2003){Garmire}, {Bautz}, {Ford}, {Nousek}, \&
  {Ricker}}]{garmire03}
{Garmire}, G.~P., {Bautz}, M.~W., {Ford}, P.~G., {Nousek}, J.~A., \& {Ricker},
  Jr., G.~R. 2003, Proc. SPIE, 4851, 28

\bibitem[{{Gelfand} \& {Gaensler}(2007)}]{gelfand07}
{Gelfand}, J.~D. \& {Gaensler}, B.~M. 2007, \apj, 667, 1111

\bibitem[{{Ghavamian} {et~al.}(2007){Ghavamian}, {Laming}, \&
  {Rakowski}}]{ghavamian07}
{Ghavamian}, P., {Laming}, J.~M., \& {Rakowski}, C.~E. 2007, \apjl, 654, L69

\bibitem[{{Gotthelf} \& {Halpern}(2005)}]{gotthelf05}
{Gotthelf}, E.~V. \& {Halpern}, J.~P. 2005, \apj, 632, 1075

\bibitem[{{Gotthelf} \& {Halpern}(2007)}]{gotthelf07}
---. 2007, \apss, 308, 79

\bibitem[{{Gotthelf} \& {Halpern}(2009)}]{gotthelf09}
---. 2009, \apjl, 695, L35

\bibitem[{{Green} {et~al.}(1999){Green}, {Cram}, {Large}, \& {Ye}}]{green99}
{Green}, A.~J., {Cram}, L.~E., {Large}, M.~I., \& {Ye}, T. 1999, \apjs, 122,
  207

\bibitem[{{Halpern} \& {Gotthelf}(2010)}]{halpern10}
{Halpern}, J.~P. \& {Gotthelf}, E.~V. 2010, \apj, 710, 941

\bibitem[{{Halpern} \& {Gotthelf}(2011)}]{halpern11}
---. 2011, \apjl, 733, L28

\bibitem[{{Halpern} {et~al.}(2005){Halpern}, {Gotthelf}, {Becker}, {Helfand},
  \& {White}}]{halpern05}
{Halpern}, J.~P., {Gotthelf}, E.~V., {Becker}, R.~H., {Helfand}, D.~J., \&
  {White}, R.~L. 2005, \apjl, 632, L29

\bibitem[{{Halpern} {et~al.}(2008){Halpern}, {Gotthelf}, {Reynolds}, {Ransom},
  \& {Camilo}}]{halpern08}
{Halpern}, J.~P., {Gotthelf}, E.~V., {Reynolds}, J., {Ransom}, S.~M., \&
  {Camilo}, F. 2008, \apj, 676, 1178

\bibitem[{{Haverkorn} {et~al.}(2006){Haverkorn}, {Gaensler},
  {McClure-Griffiths}, {Dickey}, \& {Green}}]{haverkorn06}
{Haverkorn}, M., {Gaensler}, B.~M., {McClure-Griffiths}, N.~M., {Dickey},
  J.~M., \& {Green}, A.~J. 2006, \apjs, 167, 230

\bibitem[{{Helfand} {et~al.}(2007){Helfand}, {Chatterjee}, {Brisken}, {Camilo},
  {Reynolds}, {van Kerkwijk}, {Halpern}, \& {Ransom}}]{helfand07}
{Helfand}, D.~J., {Chatterjee}, S., {Brisken}, W.~F., {Camilo}, F., {Reynolds},
  J., {van Kerkwijk}, M.~H., {Halpern}, J.~P., \& {Ransom}, S.~M. 2007, \apj,
  662, 1198

\bibitem[{{Helou} \& {Walker}(1988)}]{helou88}
{Helou}, G. \& {Walker}, D.~W., eds. 1988, {Infrared astronomical satellite
  (IRAS) catalogs and atlases. Volume 7: The small scale structure catalog},
  Vol.~7

\bibitem[{{Hong} {et~al.}(2005){Hong}, {van den Berg}, {Schlegel}, {Grindlay},
  {Koenig}, {Laycock}, \& {Zhao}}]{hong05}
{Hong}, J., {van den Berg}, M., {Schlegel}, E.~M., {Grindlay}, J.~E., {Koenig},
  X., {Laycock}, S., \& {Zhao}, P. 2005, \apj, 635, 907

\bibitem[{{Israel} {et~al.}(2010){Israel}, {Esposito}, {Rea}, {Dall'Osso},
  {Senziani}, {Romano}, {Mangano}, {G{\"o}tz}, {Zane}, {Tiengo}, {Palmer},
  {Krimm}, {Gehrels}, {Mereghetti}, {Stella}, {Turolla}, {Campana}, {Perna},
  {Angelini}, \& {de Luca}}]{israel10}
{Israel}, G.~L. \etal . 2010, \mnras, 408, 1387

\bibitem[{{Israel} \& {Stella}(1996)}]{israel96}
{Israel}, G.~L. \& {Stella}, L. 1996, \apj, 468, 369

\bibitem[{{Kalberla} {et~al.}(2005){Kalberla}, {Burton}, {Hartmann}, {Arnal},
  {Bajaja}, {Morras}, \& {P{\"o}ppel}}]{kalberla05}
{Kalberla}, P.~M.~W., {Burton}, W.~B., {Hartmann}, D., {Arnal}, E.~M.,
  {Bajaja}, E., {Morras}, R., \& {P{\"o}ppel}, W.~G.~L. 2005, \aap, 440, 775

\bibitem[{{Kaplan} {et~al.}(2009){Kaplan}, {Chatterjee}, {Hales}, {Gaensler},
  \& {Slane}}]{kaplan09}
{Kaplan}, D.~L., {Chatterjee}, S., {Hales}, C.~A., {Gaensler}, B.~M., \&
  {Slane}, P.~O. 2009, \aj, 137, 354

\bibitem[{{Kaspi}(1996)}]{kaspi96}
{Kaspi}, V.~M. 1996, in Astronomical Society of the Pacific Conference Series,
  Vol. 105, IAU Colloq. 160: Pulsars: Problems and Progress, ed. {S.~Johnston,
  M.~A.~Walker, \& M.~Bailes}, 375

\bibitem[{{Keith} {et~al.}(2010){Keith}, {Jameson}, {van Straten}, {Bailes},
  {Johnston}, {Kramer}, {Possenti}, {Bates}, {Bhat}, {Burgay}, {Burke-Spolaor},
  {D'Amico}, {Levin}, {McMahon}, {Milia}, \& {Stappers}}]{keith10}
{Keith}, M.~J. \etal . 2010, \mnras, 409, 619

\bibitem[{{Kramer} {et~al.}(2007){Kramer}, {Stappers}, {Jessner}, {Lyne}, \&
  {Jordan}}]{kramer07}
{Kramer}, M., {Stappers}, B.~W., {Jessner}, A., {Lyne}, A.~G., \& {Jordan},
  C.~A. 2007, \mnras, 377, 107

\bibitem[{{Laros} {et~al.}(1987){Laros}, {Fenimore}, {Klebesadel}, {Atteia},
  {Boer}, {Hurley}, {Niel}, {Vedrenne}, {Kane}, {Kouveliotou}, {Cline},
  {Dennis}, {Desai}, {Orwig}, {Kuznetsov}, {Sunyaev}, \& {Terekhov}}]{laros87}
{Laros}, J.~G. \etal . 1987, \apjl, 320, L111

\bibitem[{{Lazaridis} {et~al.}(2008){Lazaridis}, {Jessner}, {Kramer},
  {Stappers}, {Lyne}, {Jordan}, {Serylak}, \& {Zensus}}]{lazaridis08}
{Lazaridis}, K., {Jessner}, A., {Kramer}, M., {Stappers}, B.~W., {Lyne}, A.~G.,
  {Jordan}, C.~A., {Serylak}, M., \& {Zensus}, J.~A. 2008, \mnras, 390, 839

\bibitem[{{Levin} {et~al.}(2010){Levin}, {Bailes}, {Bates}, {Bhat}, {Burgay},
  {Burke-Spolaor}, {D'Amico}, {Johnston}, {Keith}, {Kramer}, {Milia},
  {Possenti}, {Rea}, {Stappers}, \& {van Straten}}]{levin10}
{Levin}, L. \etal . 2010, \apjl, 721, L33

\bibitem[{{Lodders}(2003)}]{lodders03}
{Lodders}, K. 2003, \apj, 591, 1220

\bibitem[{{Lorimer} {et~al.}(1995){Lorimer}, {Yates}, {Lyne}, \&
  {Gould}}]{lorimer95}
{Lorimer}, D.~R., {Yates}, J.~A., {Lyne}, A.~G., \& {Gould}, D.~M. 1995,
  \mnras, 273, 411

\bibitem[{{Manchester} {et~al.}(2001){Manchester}, {Lyne}, {Camilo}, {Bell},
  {Kaspi}, {D'Amico}, {McKay}, {Crawford}, {Stairs}, {Possenti}, {Kramer}, \&
  {Sheppard}}]{manchester01}
{Manchester}, R.~N. \etal . 2001, \mnras, 328, 17

\bibitem[{{Martini} {et~al.}(2004){Martini}, {Persson}, {Murphy}, {Birk},
  {Shectman}, {Gunnels}, \& {Koch}}]{martini04}
{Martini}, P., {Persson}, S.~E., {Murphy}, D.~C., {Birk}, C., {Shectman},
  S.~A., {Gunnels}, S.~M., \& {Koch}, E. 2004, Proc. SPIE, 5492, 1653

\bibitem[{{Mereghetti}(2008)}]{mereghetti08}
{Mereghetti}, S. 2008, \aapr, 15, 225

\bibitem[{{Monet} {et~al.}(2003){Monet}, {Levine}, {Canzian}, {Ables}, {Bird},
  {Dahn}, {Guetter}, {Harris}, {Henden}, {Leggett}, {Levison}, {Luginbuhl},
  {Martini}, {Monet}, {Munn}, {Pier}, {Rhodes}, {Riepe}, {Sell}, {Stone},
  {Vrba}, {Walker}, {Westerhout}, {Brucato}, {Reid}, {Schoening}, {Hartley},
  {Read}, \& {Tritton}}]{monet03}
{Monet}, D.~G. \etal . 2003, \aj, 125, 984

\bibitem[{{Muno} {et~al.}(2007){Muno}, {Gaensler}, {Clark}, {de Grijs},
  {Pooley}, {Stevens}, \& {Portegies Zwart}}]{muno07}
{Muno}, M.~P., {Gaensler}, B.~M., {Clark}, J.~S., {de Grijs}, R., {Pooley}, D.,
  {Stevens}, I.~R., \& {Portegies Zwart}, S.~F. 2007, \mnras, 378, L44

\bibitem[{{Murphy} {et~al.}(2007){Murphy}, {Mauch}, {Green}, {Hunstead},
  {Piestrzynska}, {Kels}, \& {Sztajer}}]{murphy07}
{Murphy}, T., {Mauch}, T., {Green}, A., {Hunstead}, R.~W., {Piestrzynska}, B.,
  {Kels}, A.~P., \& {Sztajer}, P. 2007, \mnras, 382, 382

\bibitem[{{Olausen} {et~al.}(2011){Olausen}, {Kaspi}, {Ng}, {Zhu}, {Dib},
  {Gavriil}, \& {Woods}}]{olausen11}
{Olausen}, S.~A., {Kaspi}, V.~M., {Ng}, C.-Y., {Zhu}, W.~W., {Dib}, R.,
  {Gavriil}, F.~P., \& {Woods}, P.~M. 2011, \apj, 742, 4

\bibitem[{{Osip} {et~al.}(2008){Osip}, {Floyd}, \& {Covarrubias}}]{osip08}
{Osip}, D.~J., {Floyd}, D., \& {Covarrubias}, R. 2008, Proc. SPIE, 7014, 70140A

\bibitem[{{Paladini} {et~al.}(2003){Paladini}, {Burigana}, {Davies}, {Maino},
  {Bersanelli}, {Cappellini}, {Platania}, \& {Smoot}}]{paladini03}
{Paladini}, R., {Burigana}, C., {Davies}, R.~D., {Maino}, D., {Bersanelli}, M.,
  {Cappellini}, B., {Platania}, P., \& {Smoot}, G. 2003, \aap, 397, 213

\bibitem[{{Park} {et~al.}(2012){Park}, {Hughes}, {Slane}, {Burrows}, {Lee}, \&
  {Mori}}]{park12}
{Park}, S., {Hughes}, J.~P., {Slane}, P.~O., {Burrows}, D.~N., {Lee}, J.-J., \&
  {Mori}, K. 2012, arXiv:astro-ph/1201.5056

\bibitem[{{Peretto} \& {Fuller}(2009)}]{peretto09}
{Peretto}, N. \& {Fuller}, G.~A. 2009, \aap, 505, 405

\bibitem[{{Pestalozzi} {et~al.}(2005){Pestalozzi}, {Minier}, \&
  {Booth}}]{pestalozzi05}
{Pestalozzi}, M.~R., {Minier}, V., \& {Booth}, R.~S. 2005, \aap, 432, 737

\bibitem[{{Rea} \& {Esposito}(2011)}]{rea11}
{Rea}, N. \& {Esposito}, P. 2011, in High-Energy Emission from Pulsars and
  their Systems, ed. {D.~F.~Torres \& N.~Rea} (Springer-Verlag Berlin
  Heidelberg), p. 247

\bibitem[{{Romer} {et~al.}(2001){Romer}, {Viana}, {Liddle}, \&
  {Mann}}]{romer01}
{Romer}, A.~K., {Viana}, P.~T.~P., {Liddle}, A.~R., \& {Mann}, R.~G. 2001,
  \apj, 547, 594

\bibitem[{{Scholz} \& {Kaspi}(2011)}]{scholz11}
{Scholz}, P. \& {Kaspi}, V.~M. 2011, \apj, 739, 94

\bibitem[{{Sedov}(1946{\natexlab{a}})}]{sedov46a}
{Sedov}, L.~I. 1946{\natexlab{a}}, Dokl. Akad. Nauk SSSR, 42, 17

\bibitem[{{Sedov}(1946{\natexlab{b}})}]{sedov46b}
---. 1946{\natexlab{b}}, Prikl. Mat. Mekh., 10, 241

\bibitem[{{Serylak} {et~al.}(2009){Serylak}, {Stappers}, {Weltevrede},
  {Kramer}, {Jessner}, {Lyne}, {Jordan}, {Lazaridis}, \& {Zensus}}]{serylak09}
{Serylak}, M., {Stappers}, B.~W., {Weltevrede}, P., {Kramer}, M., {Jessner},
  A., {Lyne}, A.~G., {Jordan}, C.~A., {Lazaridis}, K., \& {Zensus}, J.~A. 2009,
  \mnras, 394, 295

\bibitem[{{Skrutskie} {et~al.}(2006){Skrutskie}, {Cutri}, {Stiening},
  {Weinberg}, {Schneider}, {Carpenter}, {Beichman}, {Capps}, {Chester},
  {Elias}, {Huchra}, {Liebert}, {Lonsdale}, {Monet}, {Price}, {Seitzer},
  {Jarrett}, {Kirkpatrick}, {Gizis}, {Howard}, {Evans}, {Fowler}, {Fullmer},
  {Hurt}, {Light}, {Kopan}, {Marsh}, {McCallon}, {Tam}, {Van Dyk}, \&
  {Wheelock}}]{skrutskie06}
{Skrutskie}, M.~F. \etal . 2006, \aj, 131, 1163

\bibitem[{{Str{\"u}der} {et~al.}(2001){Str{\"u}der}, {Briel}, {Dennerl},
  {Hartmann}, {Kendziorra}, {Meidinger}, {Pfeffermann}, {Reppin}, {Aschenbach},
  {Bornemann}, {Br{\"a}uninger}, {Burkert}, {Elender}, {Freyberg}, {Haberl},
  {Hartner}, {Heuschmann}, {Hippmann}, {Kastelic}, {Kemmer}, {Kettenring},
  {Kink}, {Krause}, {M{\"u}ller}, {Oppitz}, {Pietsch}, {Popp}, {Predehl},
  {Read}, {Stephan}, {St{\"o}tter}, {Tr{\"u}mper}, {Holl}, {Kemmer}, {Soltau},
  {St{\"o}tter}, {Weber}, {Weichert}, {von Zanthier}, {Carathanassis}, {Lutz},
  {Richter}, {Solc}, {B{\"o}ttcher}, {Kuster}, {Staubert}, {Abbey}, {Holland},
  {Turner}, {Balasini}, {Bignami}, {La Palombara}, {Villa}, {Buttler},
  {Gianini}, {Lain{\'e}}, {Lumb}, \& {Dhez}}]{struder01}
{Str{\"u}der}, L. \etal . 2001, \aap, 365, L18

\bibitem[{{Sugizaki} {et~al.}(2001){Sugizaki}, {Mitsuda}, {Kaneda},
  {Matsuzaki}, {Yamauchi}, \& {Koyama}}]{sugizaki01}
{Sugizaki}, M., {Mitsuda}, K., {Kaneda}, H., {Matsuzaki}, K., {Yamauchi}, S.,
  \& {Koyama}, K. 2001, \apjs, 134, 77

\bibitem[{{Tam} {et~al.}(2006){Tam}, {Kaspi}, {Gaensler}, \&
  {Gotthelf}}]{tam06}
{Tam}, C.~R., {Kaspi}, V.~M., {Gaensler}, B.~M., \& {Gotthelf}, E.~V. 2006,
  \apj, 652, 548

\bibitem[{{Taylor}(1950{\natexlab{a}})}]{taylor50a}
{Taylor}, G. 1950{\natexlab{a}}, Royal Society of London Proceedings Series A,
  201, 159

\bibitem[{{Taylor}(1950{\natexlab{b}})}]{taylor50b}
---. 1950{\natexlab{b}}, Royal Society of London Proceedings Series A, 201, 175

\bibitem[{{Tendulkar} {et~al.}(2012){Tendulkar}, {Cameron}, \&
  {Kulkarni}}]{tendulkar12}
{Tendulkar}, S.~P., {Cameron}, P.~B., \& {Kulkarni}, S.~R. 2012, in American
  Astronomical Society Meeting Abstracts, Vol. 219, 237.03

\bibitem[{{Thompson}(2008{\natexlab{a}})}]{thompson08a}
{Thompson}, C. 2008{\natexlab{a}}, \apj, 688, 1258

\bibitem[{{Thompson}(2008{\natexlab{b}})}]{thompson08b}
---. 2008{\natexlab{b}}, \apj, 688, 499

\bibitem[{{Tiengo} {et~al.}(2010){Tiengo}, {Vianello}, {Esposito},
  {Mereghetti}, {Giuliani}, {Costantini}, {Israel}, {Stella}, {Turolla},
  {Zane}, {Rea}, {G{\"o}tz}, {Bernardini}, {Moretti}, {Romano}, {Ehle}, \&
  {Gehrels}}]{tiengo10}
{Tiengo}, A. \etal . 2010, \apj, 710, 227

\bibitem[{{Tody}(1986)}]{tody86}
{Tody}, D. 1986, in Instrumentation in astronomy VI; Proceedings of the
  Meeting, Tucson, AZ, Mar. 4-8, 1986. Part 2 (A87-36376 15-35). Bellingham,
  WA, Society of Photo-Optical Instrumentation Engineers, 1986, p. 733., ed.
  D.~L. {Crawford}, 733

\bibitem[{{Tody}(1993)}]{tody93}
{Tody}, D. 1993, in ASP Conf. Ser. 52: Astronomical Data Analysis Software and
  Systems II, ed. R.~J. {Hanisch}, R.~J.~V. {Brissenden}, \& J.~{Barnes}, 173

\bibitem[{{Turner} {et~al.}(2001){Turner}, {Abbey}, {Arnaud}, {Balasini},
  {Barbera}, {Belsole}, {Bennie}, {Bernard}, {Bignami}, {Boer}, {Briel},
  {Butler}, {Cara}, {Chabaud}, {Cole}, {Collura}, {Conte}, {Cros}, {Denby},
  {Dhez}, {Di Coco}, {Dowson}, {Ferrando}, {Ghizzardi}, {Gianotti}, {Goodall},
  {Gretton}, {Griffiths}, {Hainaut}, {Hochedez}, {Holland}, {Jourdain},
  {Kendziorra}, {Lagostina}, {Laine}, {La Palombara}, {Lortholary}, {Lumb},
  {Marty}, {Molendi}, {Pigot}, {Poindron}, {Pounds}, {Reeves}, {Reppin},
  {Rothenflug}, {Salvetat}, {Sauvageot}, {Schmitt}, {Sembay}, {Short},
  {Spragg}, {Stephen}, {Str{\"u}der}, {Tiengo}, {Trifoglio}, {Tr{\"u}mper},
  {Vercellone}, {Vigroux}, {Villa}, {Ward}, {Whitehead}, \& {Zonca}}]{turner01}
{Turner}, M.~J.~L. \etal . 2001, \aap, 365, L27

\bibitem[{{Ulmer} {et~al.}(1993){Ulmer}, {Fenimore}, {Epstein}, {Ho},
  {Klebesadel}, {Laros}, \& {Delgado}}]{ulmer93}
{Ulmer}, A., {Fenimore}, E.~E., {Epstein}, R.~I., {Ho}, C., {Klebesadel},
  R.~W., {Laros}, J.~G., \& {Delgado}, F. 1993, \apj, 418, 395

\bibitem[{{van Kerkwijk} \& {Kaplan}(2008)}]{vankerkwijk08}
{van Kerkwijk}, M.~H. \& {Kaplan}, D.~L. 2008, \apjl, 673, L163

\bibitem[{{Vasisht} \& {Gotthelf}(1997)}]{vasisht97}
{Vasisht}, G. \& {Gotthelf}, E.~V. 1997, \apjl, 486, L129

\bibitem[{{Vaughan} {et~al.}(1994){Vaughan}, {van der Klis}, {Wood}, {Norris},
  {Hertz}, {Michelson}, {van Paradijs}, {Lewin}, {Mitsuda}, \&
  {Penninx}}]{vaughan94}
{Vaughan}, B.~A. \etal . 1994, \apj, 435, 362

\bibitem[{{Yan} {et~al.}(1998){Yan}, {Sadeghpour}, \& {Dalgarno}}]{yan98}
{Yan}, M., {Sadeghpour}, H.~R., \& {Dalgarno}, A. 1998, \apj, 496, 1044

\end{thebibliography}
